\documentclass[aps]{revtex4}

\usepackage{amsmath}
\usepackage{amssymb}
\usepackage{amsfonts}
\usepackage{float}
\usepackage{graphicx}
\usepackage{pstricks}

\begin{document}
\title{An introduction to Bootstrap for nuclear physics}
\author{A. Pastore}
\affiliation{Department of Physics, University of York, Heslington, York, Y010 5DD, UK}

\begin{abstract}
This guide aims at providing a general introduction to bootstrap methods. By using simple examples taken from nuclear physics, I discuss how such a method can be used to quantify error bars of an estimator. I also investigate the use of bootstrap to estimate parameters of a simple liquid drop model. In particular, I investigate how the method compares with standard techniques based on likelihood estimator and how to take into account correlations to better evaluate confidence interval of parameters.
\end{abstract}


\date{\today}

\maketitle

\section{Introduction}

In statistics, an \emph{estimator} is a rule that applied to the sample data returns information on the underlying population. Common estimators are the mean and the variance.

%
\noindent Following Ref.~\cite{bar89}, a \emph{good} estimator should obey three criteria:  consistency, efficiency and being unbiased.
\begin{itemize}
\item Consistent: in the limit of a sample size approching the whole population, the estimator get closer and closer to the true value. 
\item Efficient: the variance of the estimator, \emph{i.e.} the fluctuations around the true value,  should be small.
\item Unbiased: for any sample size N, the estimator should on average always produce an answer close to the true value.  
\end{itemize}

\noindent For example, I can define the following estimators to determine the average height of a given population provided a sample of size $N$

\begin{itemize}
\item $\hat{a}_1$: add up all the heigths and divide by N.
\item $\hat{a}_2$: select one random observable. Ignore the rest.
\item $\hat{a}_3$:add up all the heigths and divide by N-1.
\end{itemize}

\noindent One notices that only estimator  $\hat{a}_1$ from the previous list respects all the three criteria at once. $\hat{a}_2$  is not consistent, while $\hat{a}_3$ is biased.

A common method used to define an estimator is the Maximum Likelihood Estimator (MLE). The MLE finds the set of parameters one wants to estimate that maximise the likelihood function given that specific data-set.
The estimator built from MLE is usually \emph{efficient} and \emph{consistent}, but in most cases \emph{biased}~\cite{bar89}. As a consequence it is  important to evaluate the bias and thus use it to correct predictions.

An essential ingredient for the MLE construction is to assume an underlying distribution for the data of the sample. In most cases a Gaussian assumption is reasonable, but not necessary always the most appropriate. For example, in the case of a rare radioactive decay, a Poisson distribution would be more suitable in describing data than a Gaussian.
A well-known estimator built from a Gaussian MLE is the least-square (LS) estimator. LS is widely used to determine the parameters of a given model using informations extracted from data.  
In Ref.~\cite{dob14}, the authors have illustrated in much detail all the main features of the MLE  and how to use it to derive errors.  The latter are obtained by performing a second derivative of the likelihood function with respect to a given parameter. Although the procedure looks quite simple, in reality there are some important numerical issues in how performing numerical derivatives at good level of accuracy and at reasonable computational cost~\cite{roca2015covariance}.

In this guide, I will discuss bootsrap method as a statistical tool to evaluate the bias and confidence intervals of  statistical \emph{estimators}~\cite{bar89} without the explicit use of a likelihood function.
Efron in 1979~\cite{efr79} introduced the basic idea behind bootstrap: given a data-set and a particular estimator; one builds new series of data-sets by resampling the original one. This defines a bootstrap sample.
By applying the estimator at each of these bootstrap samples, one obtains its \emph{empirical} distribution. 
The main hypothesis  is that such a distribution is very close to the \emph{true} one, so one can use it to obtain informations about the bias and the errors of the estimator. I refer to the vast available literature on the subject~\cite{efr79,efron1994introduction,dav97,chernick2011bootstrap,manly2006randomization,chernick} to demonstrate the validity of such a statement, .
The procedure I described is usually referred to as Non-Parametric Bootstrap (NPB), to differentiate it from the case where some \emph{prior} knowledge on the underlying distribution of a given estimator is given.
One defines in this case a Parametric-Bootstrap (PB): the new data-sets are randomly extracted from the postulated distribution as done in Ref.~\cite{per14} and not from the original data-set.

 Bootstrap has been rarely used in theoretical nuclear physics~\cite{Nie00,per14,bertsch2017estimating,muir2018bootstrap,pasq18}, despite its simplicity and high potential.The aim of the present guide  is to introduce the basic formalism and apply it to simple models in nuclear physics.

The guide is organised as follow: in Sec.\ref{sec:idea}, I introduce the basic idea behind bootstrap and I use it to evaluate confidence intervals of the mean and of the Pearson coefficients in two different scenario. In Sec.\ref{sec:regression}, I discuss NPB application to parameter estimates of a simple Liquid Drop model with and without explicit correlations in the data-set. I present my conclusions in Sec.\ref{sec:concl}.

\section{Bootstrap: basic concepts}\label{sec:idea}

NPB is based on the simple assumption that any experimental data-set contains informations on its parent distribution. As a consequence, if the data-set is sufficiently large, it is possible to replace the original parent distribution via the empirical one obtained via Monte Carlo methods~\cite{efr79}.

In a more precise way, let's assume I have a data-set formed by $n$ independent vectors $X=\left( x_1,x_2,\dots,x_n\right)$ and a real-valued estimator of the parameter $\hat{\theta}$. The origin of the data is not specified and thus they may be the result of a real experiment or of a simulation.
I now resample them to create a series of new data-sets named $X^*$ and I apply the estimator $\hat{\theta}^*$. 
For the sake of simplicity I consider $n=5$. A possible outcome may look as

\begin{eqnarray*}
X^*_1&=&\left( x_1,x_1,x_5,x_2,x_3\right)\;,\\
X^*_2&=&\left( x_2,x_1,x_5,x_1,x_4\right)\;,\\
X^*_3&=&\left( x_1,x_3,x_4,x_3,x_5\right)\;,\\
X^*_4&=&\left( x_2,x_2,x_2,x_2,x_5\right)\;.\\
\end{eqnarray*}

Notice that the resampling allows for repetitions.  By applying the estimator at each one of the four  bootstrap samples, $N_{\text{boot}}=4$, I obtain the experimental distribution.
From the resulting histogram, I extract confidence intervals of $\hat{\theta}$ by defining the 68\%quantile\footnote{The value of the quantile is arbitrary, other choices are possible as 90\%, 95\%\dots  } .
The resampling assumes that the observables $x_i$ are independent. Applying NPB to correlated data  destroys correlations and may lead to wrong conclusions.
I will discuss in Sec.\ref{block} an extension of the technique to take into account the case of strongly correlated data-sets.

Concerning the size of the original sample, there is a strong debate in the literature and I refer the reader to Ref.~\cite{chernick} for more details.The role of the size of the sample can be understood by considering the number of possible combinations  one can build out of a given sample

\begin{eqnarray}\label{boot:est}
\left( \begin{array}{c}2n-1\\
n \end{array}\right)=\frac{(2n-1)!}{n!(n-1)!}\;,
\end{eqnarray}

\noindent for a small sample of $n=10$, the number of possible bootstrap combinations is 92378. For $n=50$ the number of combinations grows to $\approx10^{30}$. This number need to be as big as possible so that one can consider negligible the probability of drawing two times \emph{exactly} the same sample and thus make sure all possible combinations can be extracted with equal probability.  As a rule of thumb one needs at least a sample of size $~50$, but several authors consider \emph{safe} using NPB applied to sample sizes of $\approx10-20$~\cite{fisher1990new,chernick2011bootstrap}.  
In the following, I will make use of estimator that are not sensitive to the particular order of the observables in the data-set, as for example the mean. For such a reason I used Eq.\ref{boot:est} to estimate the number of possible combinations. In the case of an estimator for which order matters, than one the number of possible sets is $n^n$.

The number of bootstrap samples should also be large to avoid introducing undesired bias. Using central limit theorem (for large samples), I assume that the difference between the empirical and real distribution follows a Gaussian distribution and the variance decreases as $\approx 1/N_{\text{boot}}$, where $N_{\text{boot}}$ is the number of bootstrap samplings.
In the present article, due to the low computational cost of Monte-Carlo sampling, I  will use typically $N_{\text{boot}}\approx10^4$ and thus neglect the bias.

Before moving forward, it is important mentioning a tecnique very similar to NPB and named \emph{jackknife}~\cite{mil74}. In this case instead of resampling the full data-set one leaves out $m$ observables at the time   and then one   uses the remaining $N-m$ data  to evaluate the estimator. Although it looks very similar to NPB, the mathematical hypothesis behind it are very different~\cite{tuk58}.
Jackknife is computationally less expensive than NPB, but it is specialised to obtain the variance of the estimator, while NPB gives access to the complete empirical distribution.

\subsection{Example: confidence interval of the mean}

I illustrate the power of bootstrap by applying it to the well-known case of the estimator of the mean $\hat{\mu}$ of a data-set $\{X_1,X_2,\dots,X_N\}$ of length $N$.
In this example, I  will consider a  sample of size $N=10$ extracted from an exponential distribution with mean $\frac{1}{\lambda}=\frac{1}{2}$. In Tab.\ref{tab:data}, I report the values used for this example.

\begin{table}[!h]
\begin{center}
\begin{tabular}{c|cccccccccc}
\hline
\hline
value &0.068& 1.649& 0.058& 0.165 & 0.522& 0.040 & 1.078& 0.512 & 0.354& 0.449 \\
 position &1       & 2       & 3           &  4 & 5& 6&7 &8 &9 &10 \\
\hline
\hline
\end{tabular}
\label{tab:data}
\caption{Random values extracted from exponential distribution with mean $\frac{1}{\lambda}=\frac{1}{2}$.}
\end{center}
\end{table}

\noindent 
To calculate the mean of the parent distribution, I use the estimator

\begin{eqnarray}\label{eq:est}
\hat{\mu}=\frac{1}{N}\sum_{i=1}^NX_i\;.
\end{eqnarray}

In this particular case I obtain $\hat{\mu}=0.489$. The estimator is \emph{efficient}, \emph{consistent} and \emph{unbiased}~\cite{bar89}, moreover for this particular case (knowing the distribution of the estimator) it is possible to calculate the resulting error bars using the formula
\begin{eqnarray}
\sigma_M=\frac{\sigma}{\sqrt{N}}\;,
\end{eqnarray}
where $\sigma$ is the root mean square of the data-set.
I obtain $\sigma_M=0.154$. 

\begin{figure*}[!h]
\begin{center}
\includegraphics[width=0.35\textwidth,angle=-90]{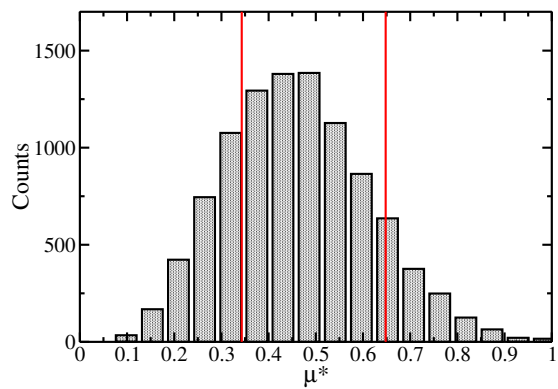}
\end{center}
\caption{(Colors online) Confidence interval of the mean: distribution of occurrences of the value $\mu^*$ for the mean, Eq. \ref{eq:est}, obtained from various NPB samples. The vertical lines indicate the confidence interval defining 68\% of the total counts.}
\label{fig:mean}
\end{figure*}

I  now repeat the same analysis using NPB. In this case I perform a random sampling ($N_{boot}=10^4$) and  I apply the estimator $\mu^*$.
I group the results using an histogram as illustrated in Fig.\ref{fig:mean}. On this figure, I also show the confidence interval of the NPB estimates by considering the lower (upper) quantiles defining 68\% of the total counts.

The average value of the different means calculated  via NPB is $\bar{\mu^*}=0.489_{-0.146}^{0.159}$. As expected the values coincide since, it is known that the estimator in Eq.\ref{eq:est} has no bias. The error bars are not symmetric (since I assume no prior Gaussian distribution), but they are extremely close to the estimated value, thus showing that NPB is a reliable method.

\subsection{Example: confidence interval on Pearson coefficient}

In this section, I consider another estimator, \emph{i.e.} the Pearson coefficient~\cite{bar89}. The standard formulas to estimate errors of the Pearson coefficient are based on the Fisher transformation~\cite{cor98}. 
The latter is valid under the assumption of normally distributed data, so in case of outliers it is possible to obtain inconsistent results. 

As an example, I consider the possible correlation between the slope of the symmetry energy $L_0$ and the neutron skin thickness $\Delta r_{np}$ of heavy nuclei~\cite{cen09,roc11}.

\begin{figure*}[!h]
\begin{center}
\includegraphics[width=0.45\textwidth,angle=-90]{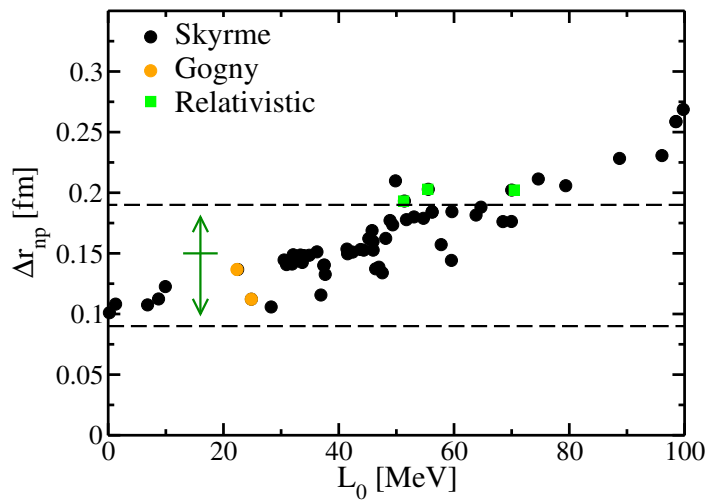}
\end{center}
\caption{(Colors online)  Neutron skin thickness in $^{208}$Pb as a function of $L_0$ for various nuclear models. The horizontal dashed line represent the experimental confidence interval as taken from Ref.~\cite{tar14}. See text for details.}
\label{fig:skin}
\end{figure*}

In Fig.\ref{fig:skin}, I show the neutron skin thickness in $^{208}$Pb as a function of $L_0$ extracted from several nuclear models. The data-set is composed by 63 nuclear functionals grouped in 3 big families: 
zero-range Skyrme functionals~\cite{per04}; finite-range Gogny~\cite{dec80} and  Relativistic Lagrangians~\cite{ben03,nik08} . For completeness, I report the models used in the calculation in Tab.\ref{tab:inm} with all relevant infinite nuclear matter properties. 
 
The neutron skin thickness is extracted from the full quantal calculation using two point Fermi function (2pF) as detailed in Ref.~\cite{war10}.
The two horizontal lines represent the region compatible with the most recent experimental  measurement of neutron skin thickness in $^{208}$Pb~\cite{tar14} with its corresponding experimental error bars.
To study the possible correlation between $L_0$  and $\Delta r_{np}$, I calculate the Pearson coefficient as

\begin{eqnarray}\label{eq:pear}
\hat{r}=\frac{\sum_{i=1}^N(X_i-\bar{X})(Y_i-\bar{Y})}{\sigma_X\sigma_Y}
\end{eqnarray}

\noindent where $\bar{X},\bar{Y}$ represent the mean values and $\sigma_{X,Y}$ the root mean square of the sample.

\begin{figure}[!h]
\begin{center}
\includegraphics[width=0.38\textwidth,angle=-90]{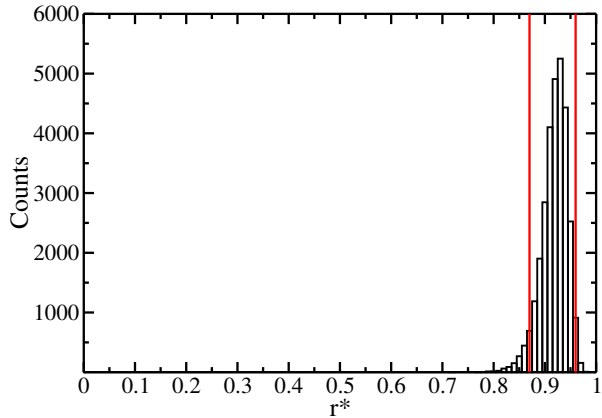}
\end{center}
\caption{(Colors online) Confidence interval of the Pearson coefficient: distribution of occurrences of the values of  $r^*$, Eq.\ref{eq:pear}, obtained using various bootstrap data sets in $^{208}$Pb. The vertical lines represent the region including 95\% of the counts. }
\label{boot1}
\end{figure}

\noindent For the present case I obtain $\hat{r}=0.91^{0.03}_{-0.05}$ where the error bars represent  95\% confidence interval obtained using Fisher transformation.
Performing NPB on the same data, I obtain a distribution of Pearson coefficient $r^*$ as shown in Fig.\ref{boot1}. The vertical lines represent the region including the 95\% counts.

The average value of the correlation coefficient extracted from NPB is

\begin{eqnarray}
\bar{r}^*=0.92^{0.04}_{-0.04} \;.
\end{eqnarray}

The results are in excellent agreement between the two methods and with Refs~\cite{cen09,roc11}. I conclude that the correlation in this data-set is very robust and very little dependent on the particular choice of the data used. This can be understood by looking at the distribution in Fig.\ref{boot1} and keeping in mind that \emph{not} all data are used in every bootstrap resampling. 

I repeat the previous analysis, but considering a different data-set to evaluate the impact of possible outliers when extracting statistical informations.

\begin{figure*}[!h]
\begin{center}
\includegraphics[width=0.45\textwidth,angle=-90]{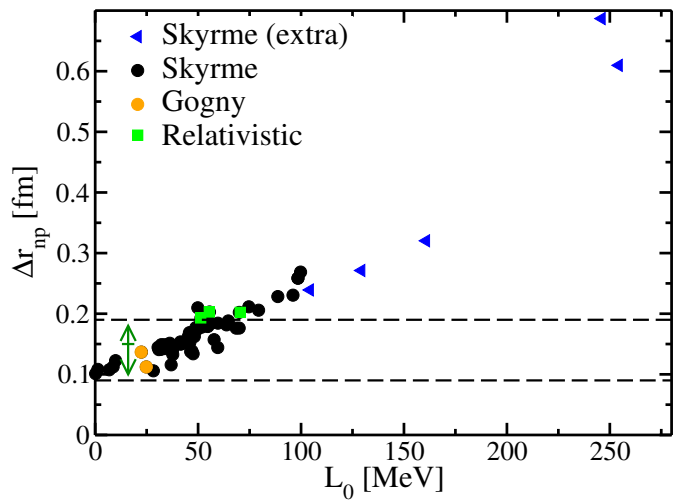}

\end{center}
\caption{(Colors online)  Same as Fig.\ref{fig:skin}, but with 5 additional Skyrme functional (triangles). }
\label{fig:skin2b}
\end{figure*}

In Fig.\ref{fig:skin2b}, I show the neutron skin thickness of $^{208}$Pb calculated with the same data-set reported in Tab.\ref{tab:inm}, with an addition of extra 5 points (triangles).
These points correspond to  5 Skyrme functionals with very large values of $L_0$: SEFM68~\cite{dut12}, SGOII ~\cite{she09}, SKI1-2-5~\cite{reinhard1995nuclear}.
These functionals produce a value of $L_0$ which is  way too large compared to the acceptable values of $L_0$~\cite{dut12}.
In this guide, I assume for simplicity that each observables can be resambles with the same probability. Given the \emph{prior} knowledge on the acceptable values of $L_0$~\cite{dut12}, one may use instead the Bayesian version of bootstrap. I refer the reader to Refs. \cite{rubin1981bayesian,boos1986bootstrap} for more details on the method.

I performed the same NPB analysis as before in $^{208}$Pb, but for this extended data-set. I obtain from NPB the value $\bar{r}^*=0.96^{0.03}_{-0.03}$. The distribution of $r^*$ reseambles very much the one reported in Fig.\ref{boot1}, although much more narrow. The extra points do not  change the conclusion of the previous section.

\subsection{Example: confidence interval on Pearson coefficient with outliers}

I now consider a different close shell nucleus: $^{100}$Sn. In a previous study~\cite{muir2018bootstrap}, the authors have shown that there is no real correlation within $\Delta r_{np}$ and $L_0$ in the case of a close shell nucleus with equal number of neutrons and protons, since the possible presence of a proton skin is mainly driven by Coulomb interaction more than by symmetry energy properties.

\begin{figure*}[!h]
\begin{center}
\includegraphics[width=0.45\textwidth,angle=-90]{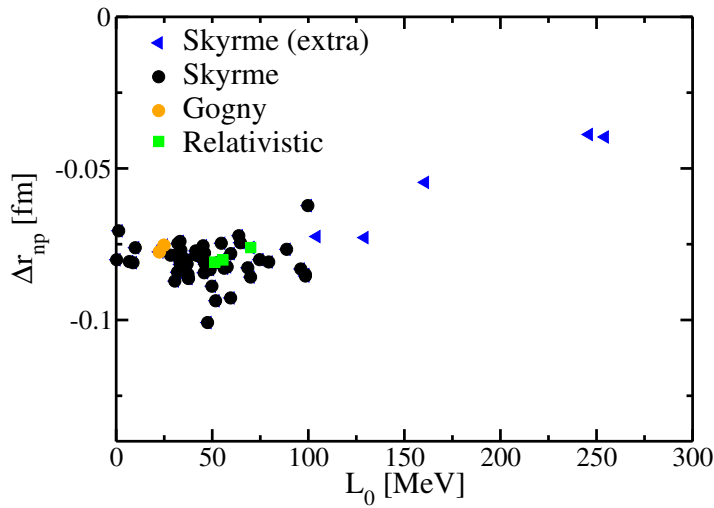}
\end{center}
\caption{(Colors online)  Neutron skin thickness in $^{100}$Sn as a function of $L_0$ for various models. See text for details.}
\label{fig:skin2}
\end{figure*}

This case is quite suitable to illustrate how the presence of outliers may lead to wrong conclusions when using the Pearson coefficient and the Fisher transformation to estimate error bars.
In Fig.\ref{fig:skin2}, I show the neutron skin thickness of $^{100}$Sn calculated with the same data-set reported in Tab.\ref{tab:inm}, with an addition of extra 5 points (triangles). I use the same data set used to produce  Fig.\ref{fig:skin2b}.

The Pearson coefficient calculated with these extra 5 Skyrme functionals leads to $\hat{r}=0.68^{0.11}_{-0.15}$. The correlation coefficient is smaller than  the one obtained in $^{208}$Pb, but I can exclude with statistical significance the null-hypothesis of a non-correlated set.

In this case NPB may be very useful due to its random resampling nature: if  the data are contaminated by few outliers, during random sampling most likely one will produce new data-sets with no outliers and thus getting strong variation on the estimator. Similar results would  also be obtained using a much simpler  \emph{jackknife}  method as illustrated in Ref.~\cite{efr79,efron1981jackknife}.
Actually this is a quite common procedure used to validate how the conclusions of a statistical analysis depend on the particular choice of the data sample.

\begin{figure}[!h]
\begin{center}
\includegraphics[width=0.38\textwidth,angle=-90]{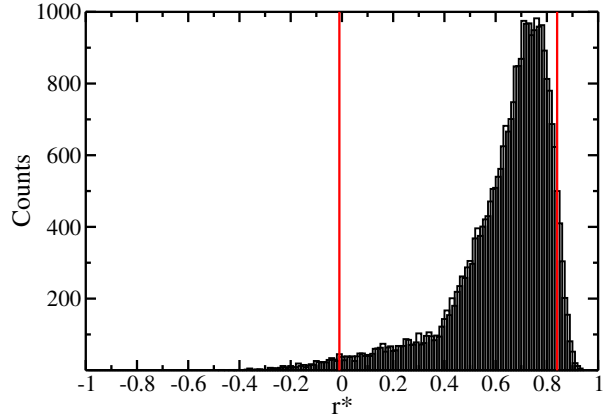}
\end{center}
\caption{(Colors online)  Confidence interval of the Pearson coefficient: distribution of occurrences of the values of  $r^*$, Eq.\ref{eq:pear}, obtained using various bootstrap data sets in $^{100}$Sn. The vertical lines represent the region including 95\% of the counts.}
\label{boot2}
\end{figure}

In Fig.\ref{boot2}, I report the empirical distribution of the Pearson estimators $r^*$ calculated using $N_{boot}=3\times10^4$ Monte-Carlo samplings. Contrary to Fig.\ref{boot1}, the distribution appears to be much more broad. The calculated  mean value with 95\% confidence error bar is $\bar{r}^*=0.58^{0.25}_{-0.59}$. Although the mean value of the estimator obtained with bootstrap is in good agreement with the direct calculation of the Pearson coefficient on the original the data-set, I observe that the error bars are much larger. As a consequence it is not possible to exclude the null-hypothesis of non-correlated data with a good statistical significance.

In this example, NPB is not used to identify which point is an outlier, but inspecting Fig.\ref{boot2}, I can clearly observe a very large variance in the resulting estimator depending on the very specific choice of the observables of the data-set.  This outcome should be used as a signal for a further investigation of the observables contained in  the data-set.

In this specific case, I decided to exclude the  extra 5 points as done in Ref.~\cite{muir2018bootstrap}, by advokating the criteria on $L_0$ given in Ref~\cite{dut12}. I obtain a result with a much more narrow distribution and centred on $r^*=0\pm0.3$. It is important to stress the crucial role of error bars in this example. In both cases (with and without the 5 extra points) although the mean Pearson coefficient is very different, the error bars obtained via NPB lead always to the same conclusion: one can not exclude the null-hypothesis of non-correlated data.


\section{Regression analysis}\label{sec:regression}

Most of the models currently applied in theoretical nuclear physics rely on some adjustable parameters, usually obtained via LS minimisation~\cite{dob14}.
In more general terms, the model function $M(\mathbf{x},\mathbf{p})$ depends on some input variables $\mathbf{x}$ and parameters $\mathbf{p}$. This function is derived under some assumptions and it contains our knowledge of a particular (physical) phenomenon.
The least-square estimator is used to determine the optimal set of parameters $\mathbf{p}_0$ to maximise (minimise) the likelihood function of the parameters assuming an underlying Gaussian distribution.

Given the observable $Y(\mathbf{x})$, I have the following equation

\begin{equation}\label{ed:mod}
Y(\mathbf{x})=M(\mathbf{x},\mathbf{p}_0)+\mathcal{E}(\mathbf{x})\;,
\end{equation}

\noindent where $\mathcal{E}(\mathbf{x})$ represents a residual error.  The standard assumption is that the residuals are uncorrelated and they follow a normal distribution with zero mean. 


In this article, I discuss how to use  NPB  to estimate the parameters  $\mathbf{p}_0$  and their error bars in both cases of non-correlated and strongly correlated residuals.
The  main advantage of NPB compared to the method presented in Ref.~\cite{dob14} is that  having access to the empirical distribution, I can extract error bars on parameters directly from the histogram without performing  numerical derivatives in parameter space~\cite{roca2015covariance,robin}.

Since the present article aims at illustrating the methodology, I will consider the simple liquid drop (LD) model to calculate binding energies, $B$, for various nuclei as a function of neutron ($N$) and proton ($Z$) number.
The binding energy is calculated as

\begin{eqnarray}\label{bene}
B_{th}(N,Z)=a_v A-a_sA^{2/3}-a_c\frac{Z(Z-1)}{A^{1/3}}-a_a\frac{(N-Z)^2}{A}-\delta\frac{\text{mod}(Z,2)+\text{mod}(N,2)-1}{A^{1/2}}\;,
\end{eqnarray}

\noindent where $A=N+Z$. The set of parameters I want  to determine is $\mathbf{p}=\{a_v,a_s,a_c,a_a,\delta \}$. These parameters are  named volume ($a_v$), surface ($a_s$), Coulomb ($a_c$), asymmetry ($a_a$) and pairing term ($\delta$) and they refer to specific physical properties of the underlying nuclear system. I refer to Ref.~\cite{krane1988introductory}, for  a more detailed discussion on the physical meaning of these parameters.
This model has a linear dependence on the parameters, so in principle, it is possible to perform least-square minimisation analytically by simple matrix inversion~\cite{bar89}. I will not use such a specific feature of the model, since I want to keep the discussion as general as possible.
The experimental nuclear binding energies ,$B_{exp}(N,Z)$, are extracted from Ref.~\cite{wang2012ame2012}. In my calculation I exclude all nuclei with $A<14$, since Eq.\ref{bene} is not suitable to describe very light systems. I define the penalty function as

\begin{eqnarray}\label{chi2}
\chi^2=\sum_{N,Z\in\text{data-set}}\frac{(B_{exp}(N,Z)-B_{th}(N,Z))^2}{\sigma^2(N,Z)}\;.
\end{eqnarray}

Moreover, I will consider in my calculation only measured binding energies  with an experimental error lower than $100$ keV. So  in first approximation I set $\sigma^2(N,Z)=1$, thus assuming equal weight to all data.

\begin{figure}[!h]
\begin{center}
\includegraphics[width=0.38\textwidth,angle=-90]{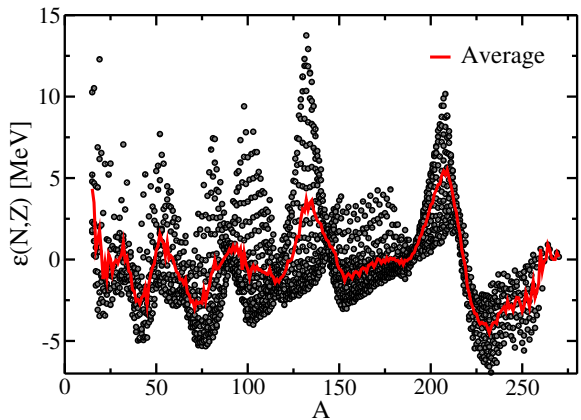}
\end{center}
\caption{(Colors online) Residuals $\mathcal{E}(N,Z)$ for LD model obtained using least square minimisation. The solid line represents the average residual for fixed values of A as defined in Eq.\ref{av:res}.}
\label{resid}
\end{figure}

The LS  fit provides the values reported in Tab.\ref{tab:fit}. Following Refs.~\cite{dob14,perez2015error}, the penaly function given in Eq.\ref{chi2} normalised by the number of degrees of freedom  should be equal to one  at the minimum. This is not the case here, I have thus introduced a global scaling constant, usually named Birge factor \cite{klupfel2009variations},  corresponding to the root mean square deviation of the residuals presented in Fig.\ref{resid}.
The errors are obtained by taking the square root of the diagonal part of the variance matrix.

\begin{table}
\begin{center}
\begin{tabular}{c|cc}
\hline
\hline
Parameter             &  [MeV] &  Error [MeV]\\
             \hline
 $a_v$&   15.69  &       $\pm$0.025\\
  $a_s$&   17.75 &        $\pm$0.08\\
 $a_c$&  0.713 &        $\pm$0.002\\
 $a_a$   &23.16  &      $\pm$0.06\\
  $\delta$&   11.8&       $\pm$0.9 \\
\hline
\hline
\end{tabular}
\caption{Parameters  of LD model obtained using least square fitting. On the last column I report the error on the fitted parameters.}
\label{tab:fit}
\end{center}
\end{table}

\begin{table}
\begin{center}
\begin{tabular}{c|ccccc}
\hline
\hline
&$a_v$ & $a_s$ &$a_c$ & $a_a$ & $\delta$\\
\hline 
 $a_v$& 1 &        & & & \\
$a_s$  &0.993 &       1 &       & & \\
$a_c$  &0.984 &       0.960 &       1 &      &\\
$a_a$  &0.913 &      0.897 &       0.879 &        1&\\
$\delta$&0.036 & 0.035 &  0.038 &  0.036&   1\\  
\hline
\hline
\end{tabular}
\caption{Correlation matrix for LD parameters obtained from LS fitting.}
\label{tab:corr}
\end{center}
\end{table}

As already stated in Ref.~\cite{bertsch2017estimating}, the errors obtained from LS under the assumption of non-correlated observables are quite unrealistic compared to the structure of the residuals $\mathcal{E}(N,Z)$ shown in Fig.\ref{resid}.

The variance of the residuals is quite large ($\approx$ few MeV).  I tested that the null hypothesis that the residuals are distributed according to a normal distribution is rejected at the 5\% level based on the Kolmogorov-Smirnov test ~\cite{bar89}. 
 This means that there is a strong signal left in the residuals and Eq.\ref{ed:mod} is not valid. In this case, I conclude that the model given in Eq.\ref{bene} is not satisfactory and extra terms should be included to improve the fit \cite{pomorski2003nuclear}.

%
%

\subsection{Uncorrelated Bootstrap}

\begin{figure}[h]
\begin{center}$
\begin{array}{ccccc}
\includegraphics[width=0.18\textwidth,angle=0]{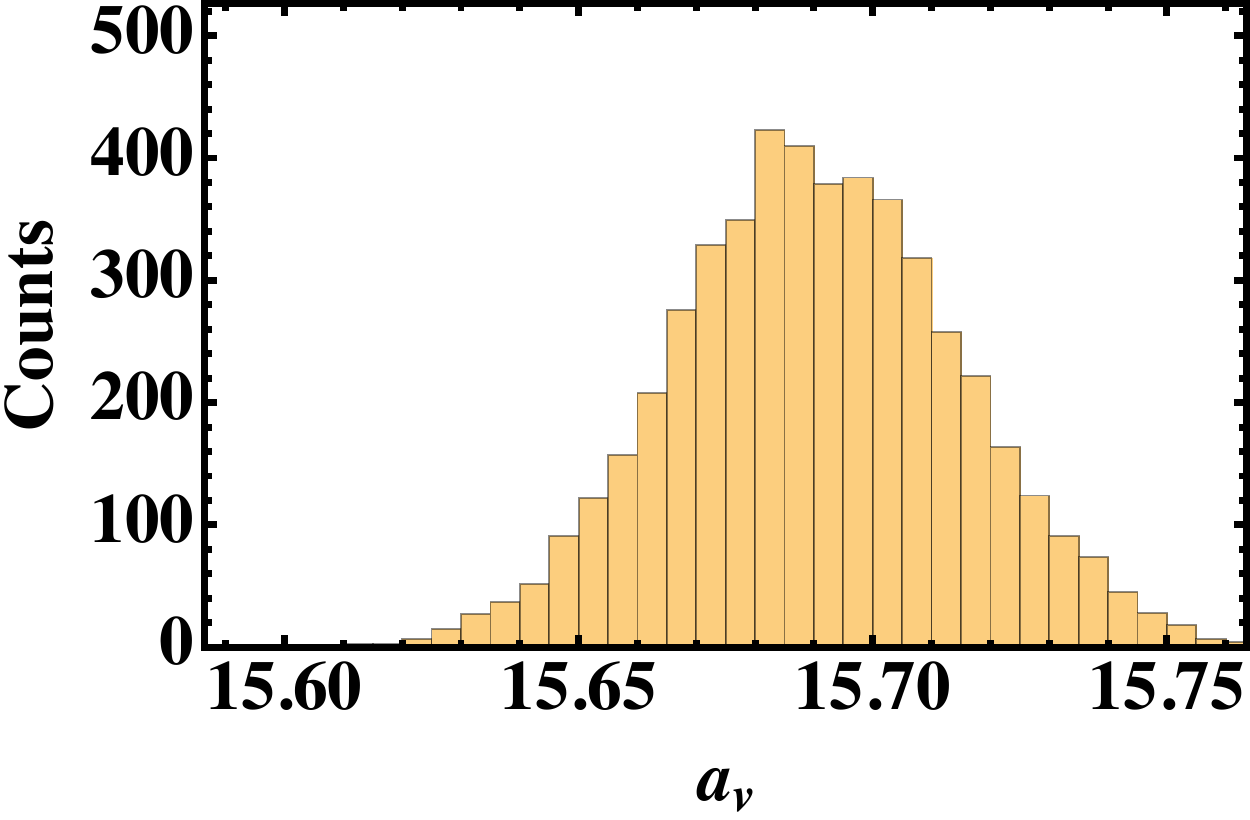} & &  && \\
\includegraphics[width=0.18\textwidth,angle=0]{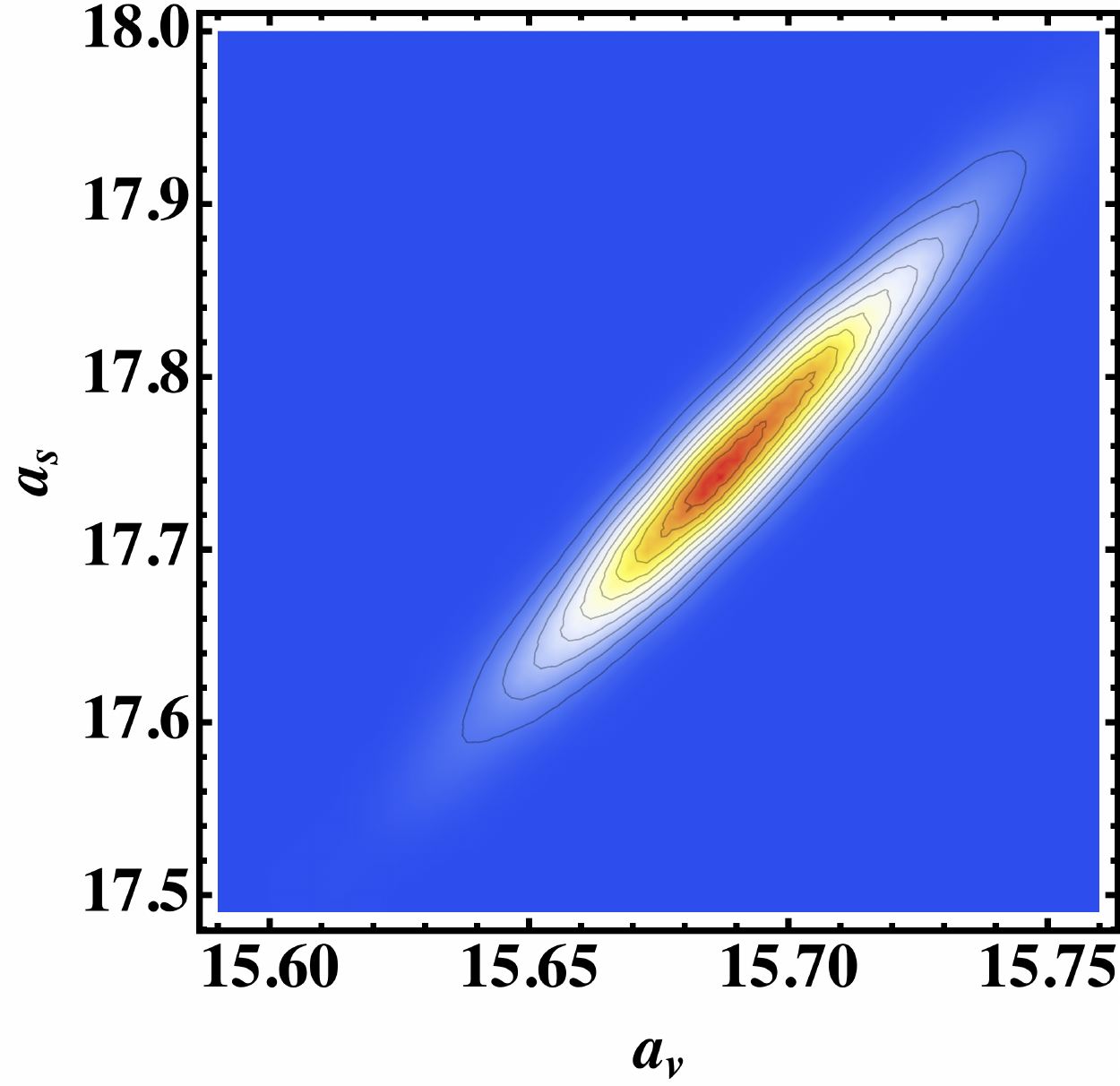} & \includegraphics[width=0.18\textwidth,angle=0]{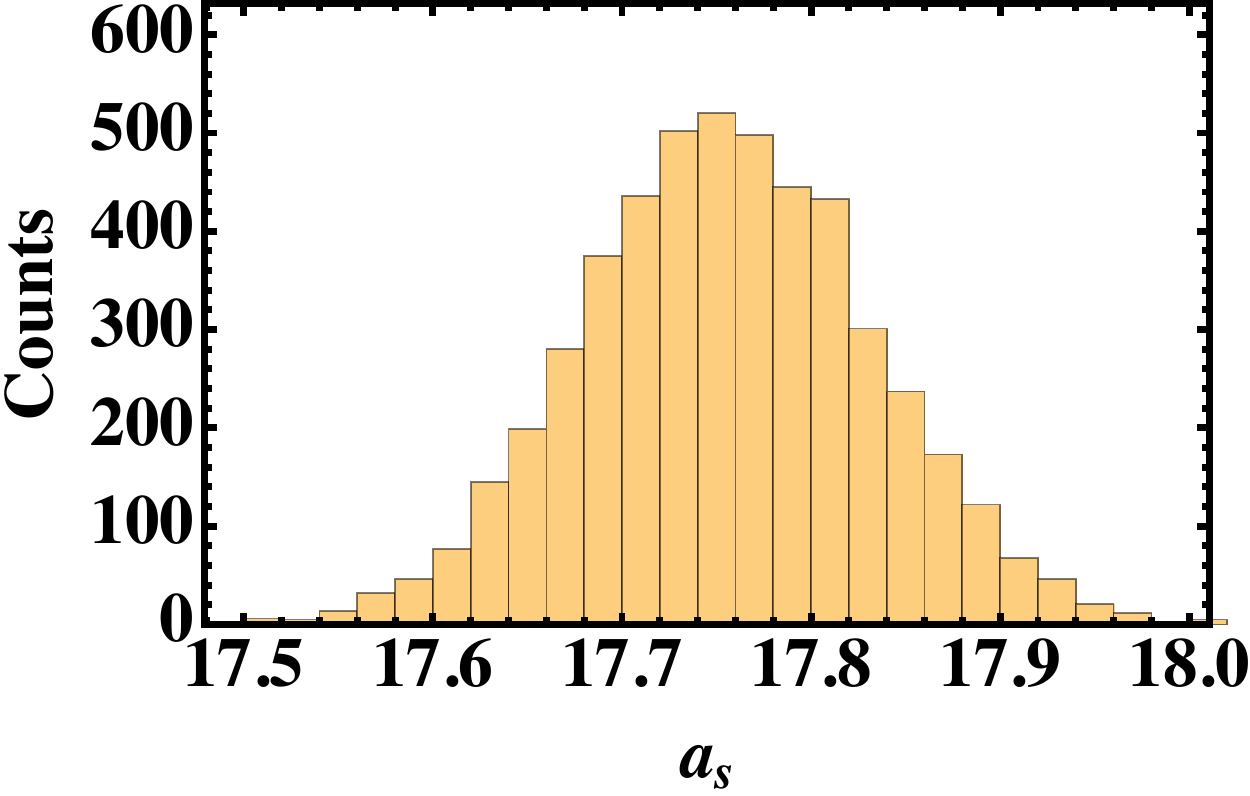}&  & &\\
\includegraphics[width=0.18\textwidth,angle=0]{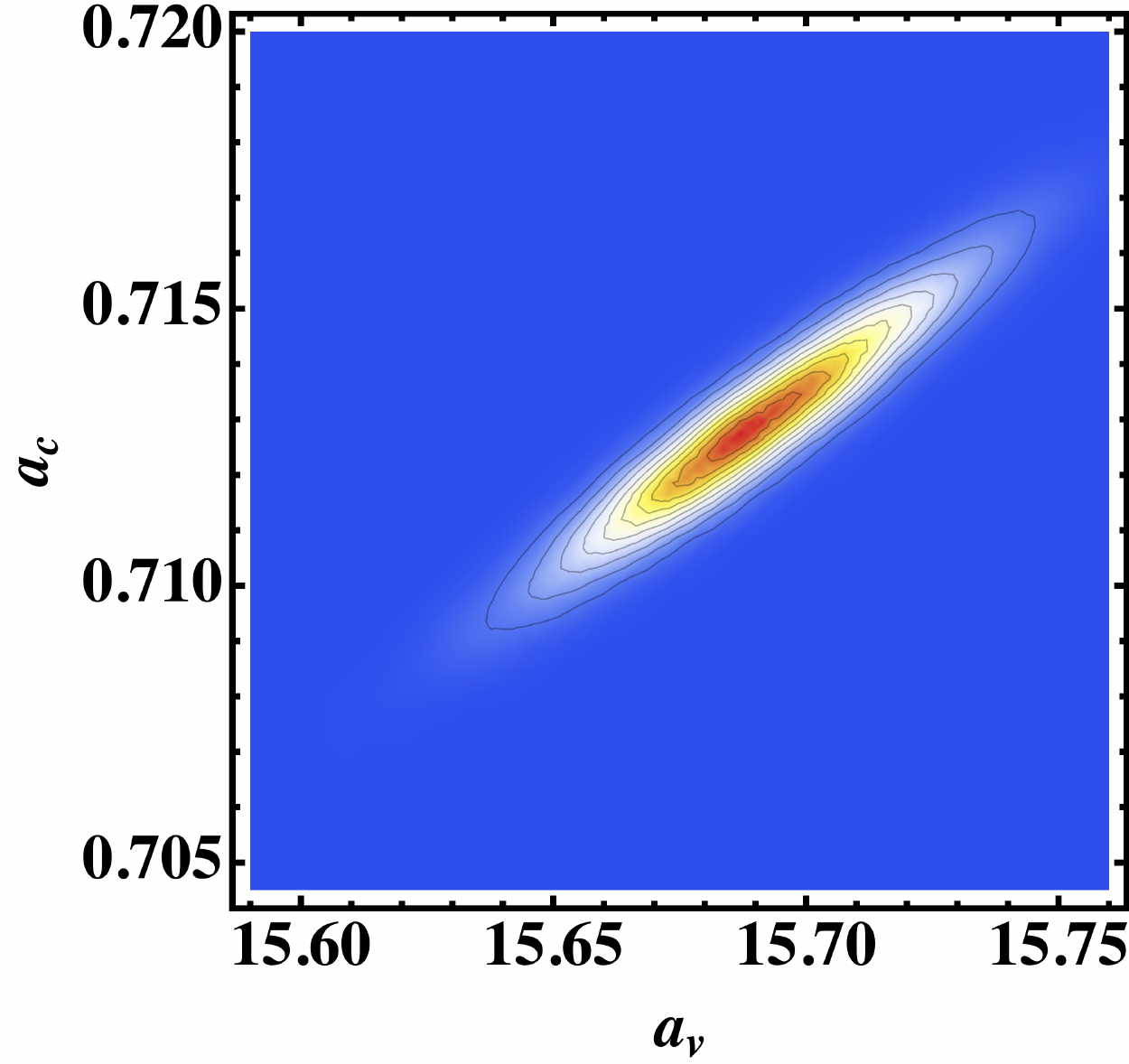} &\includegraphics[width=0.18\textwidth,angle=0]{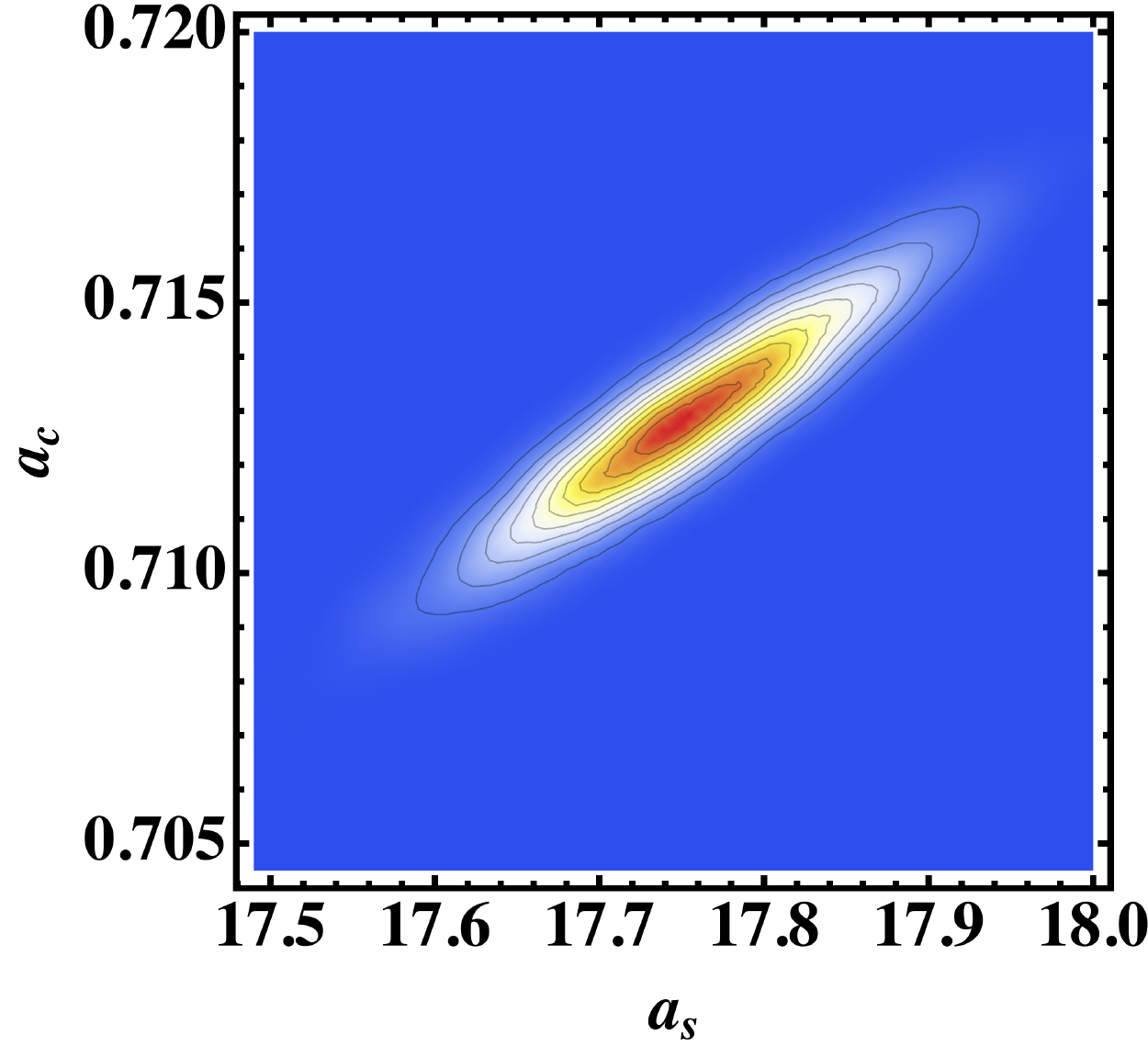}& \includegraphics[width=0.18\textwidth,angle=0]{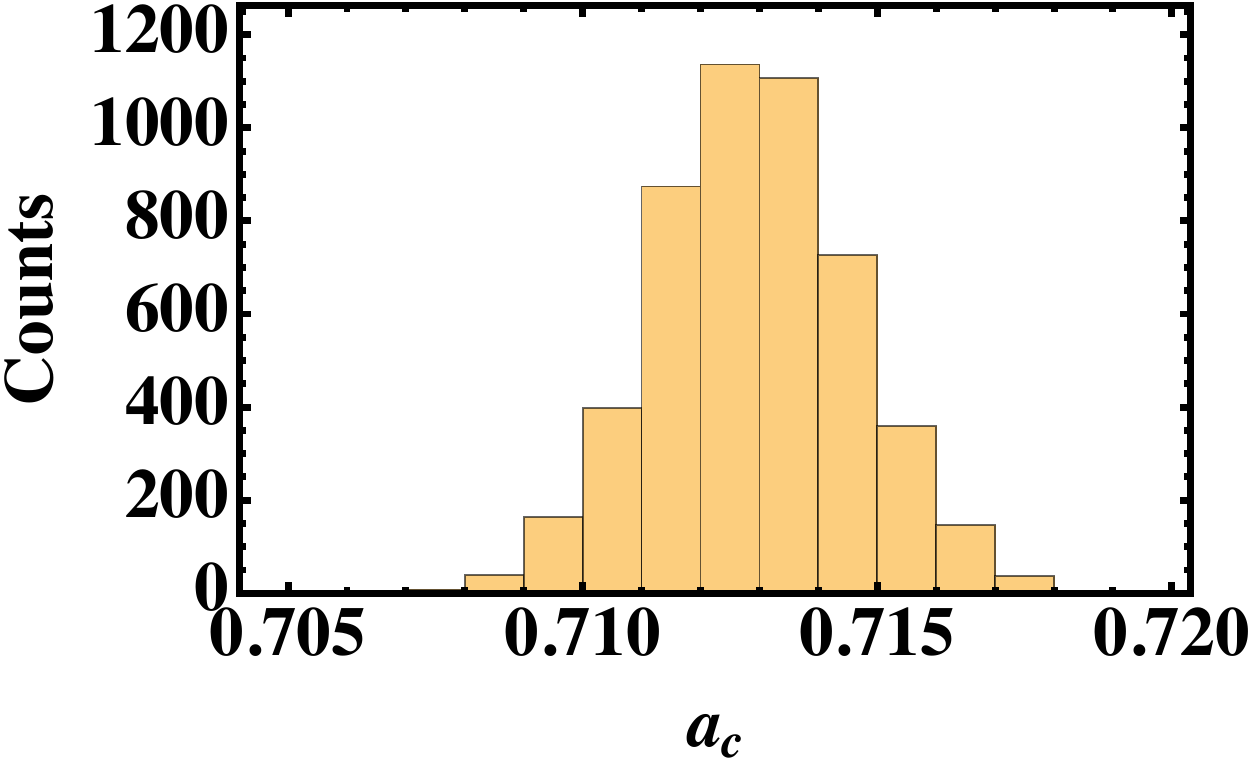}&& \\ 
\includegraphics[width=0.18\textwidth,angle=0]{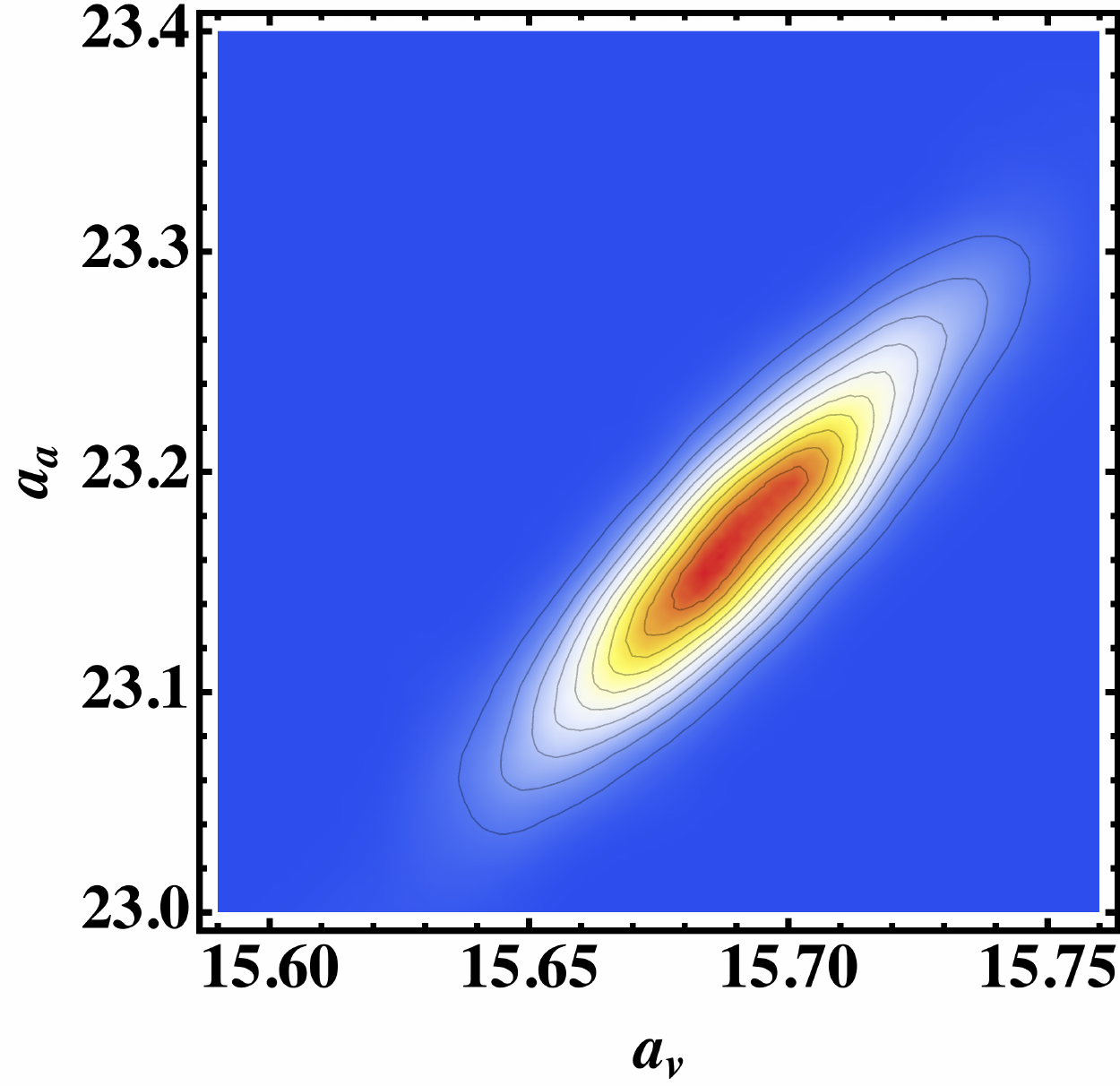} &\includegraphics[width=0.18\textwidth,angle=0]{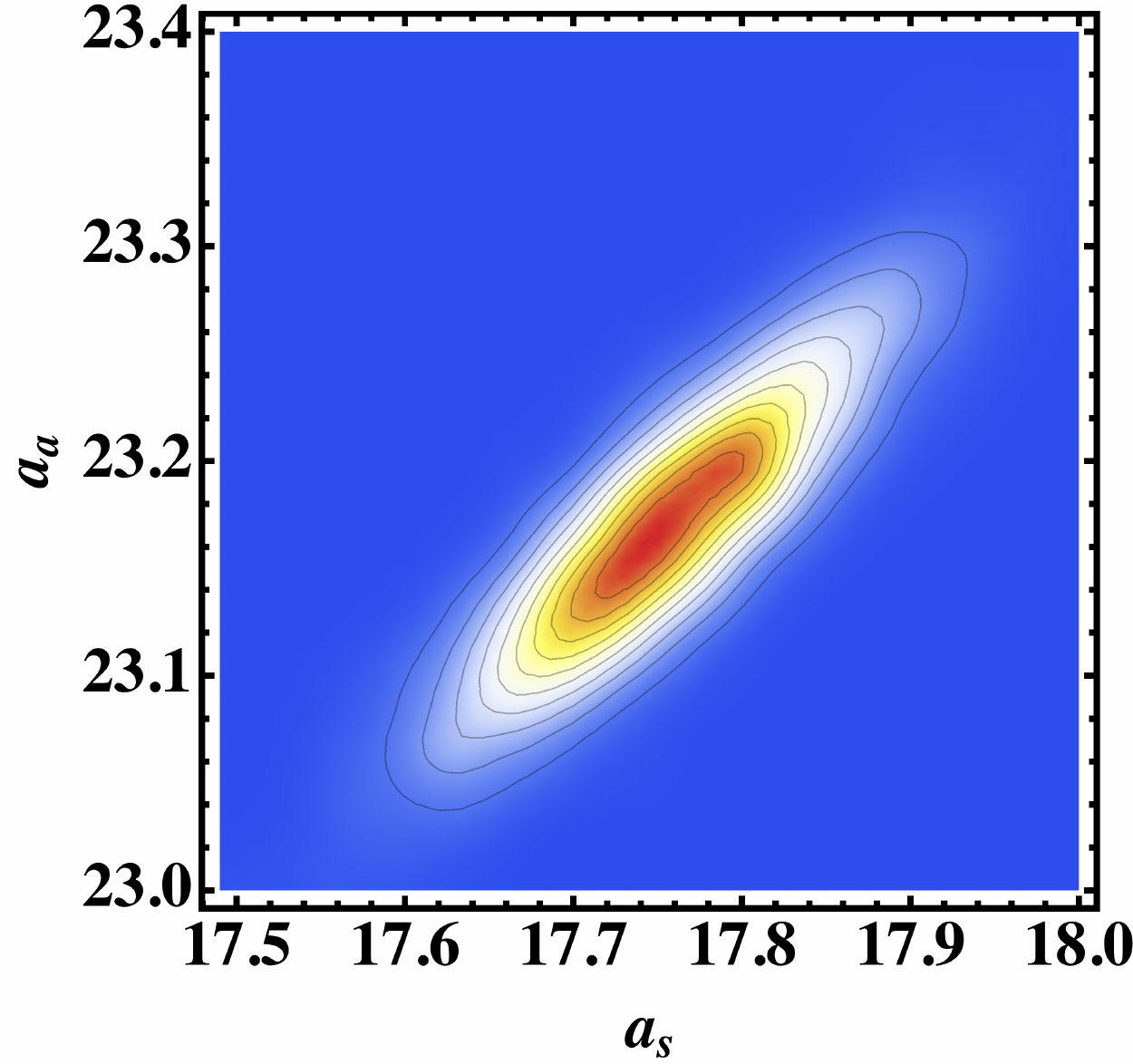}&\includegraphics[width=0.18\textwidth,angle=0]{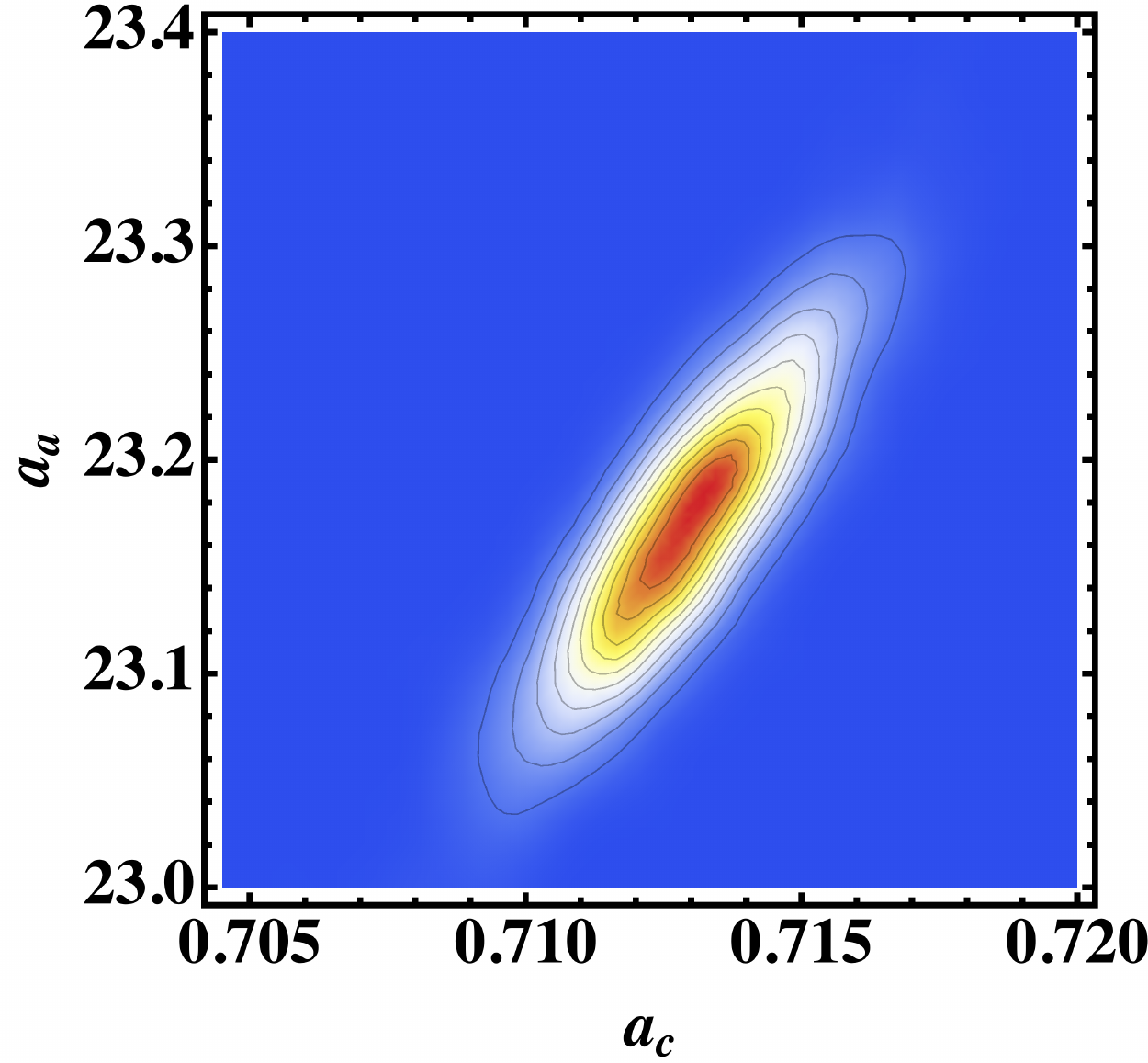}&\includegraphics[width=0.18\textwidth,angle=0]{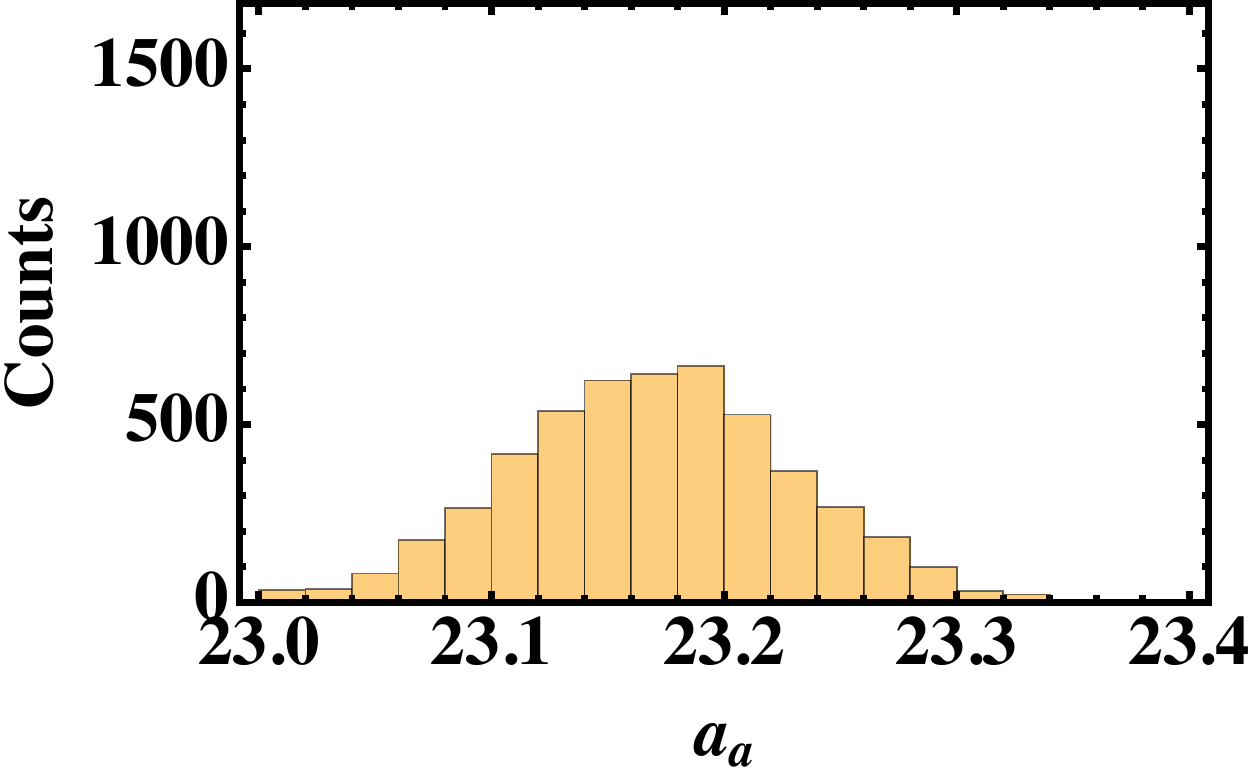} &\\ 
\includegraphics[width=0.18\textwidth,angle=0]{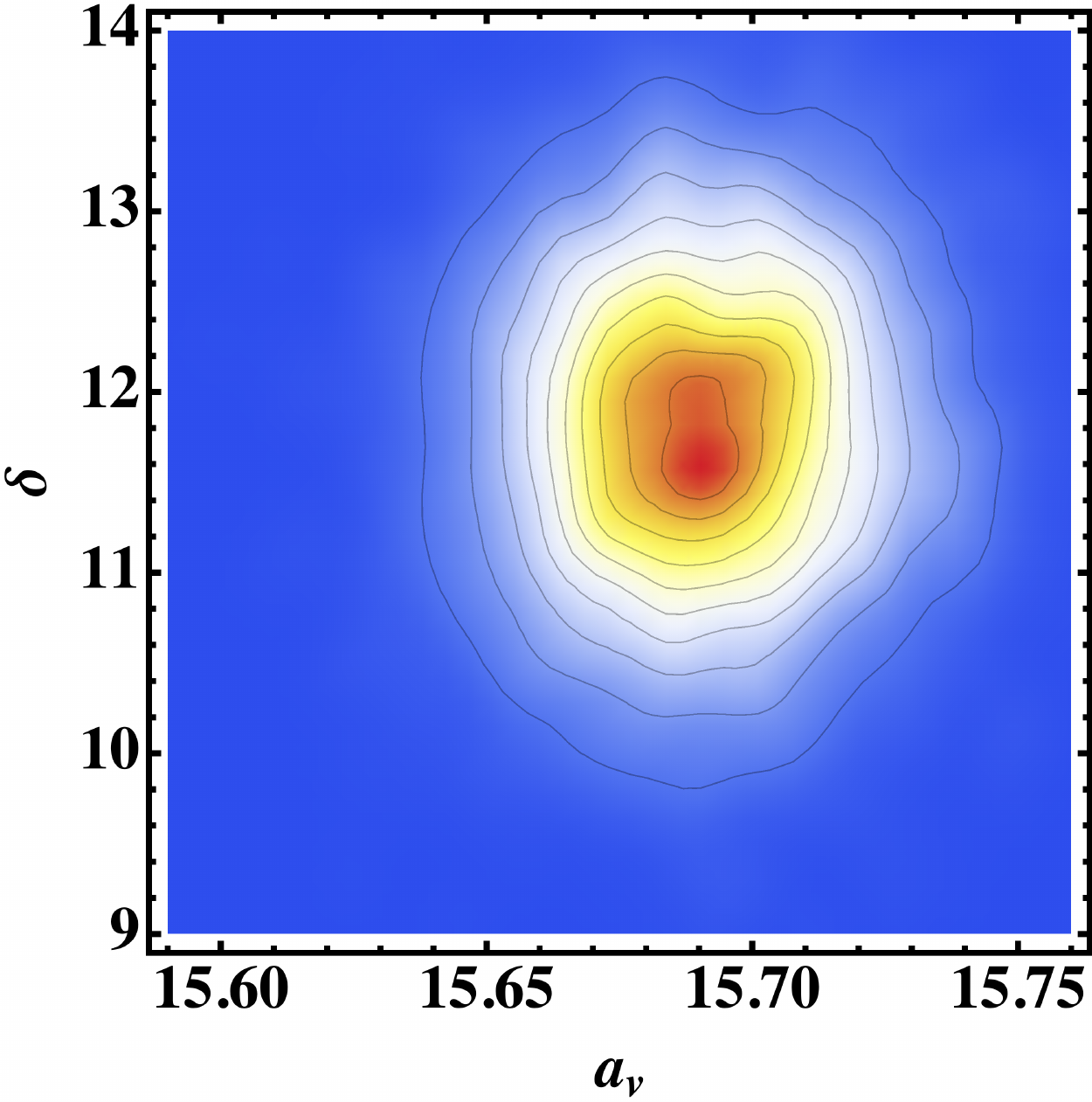} &\includegraphics[width=0.18\textwidth,angle=0]{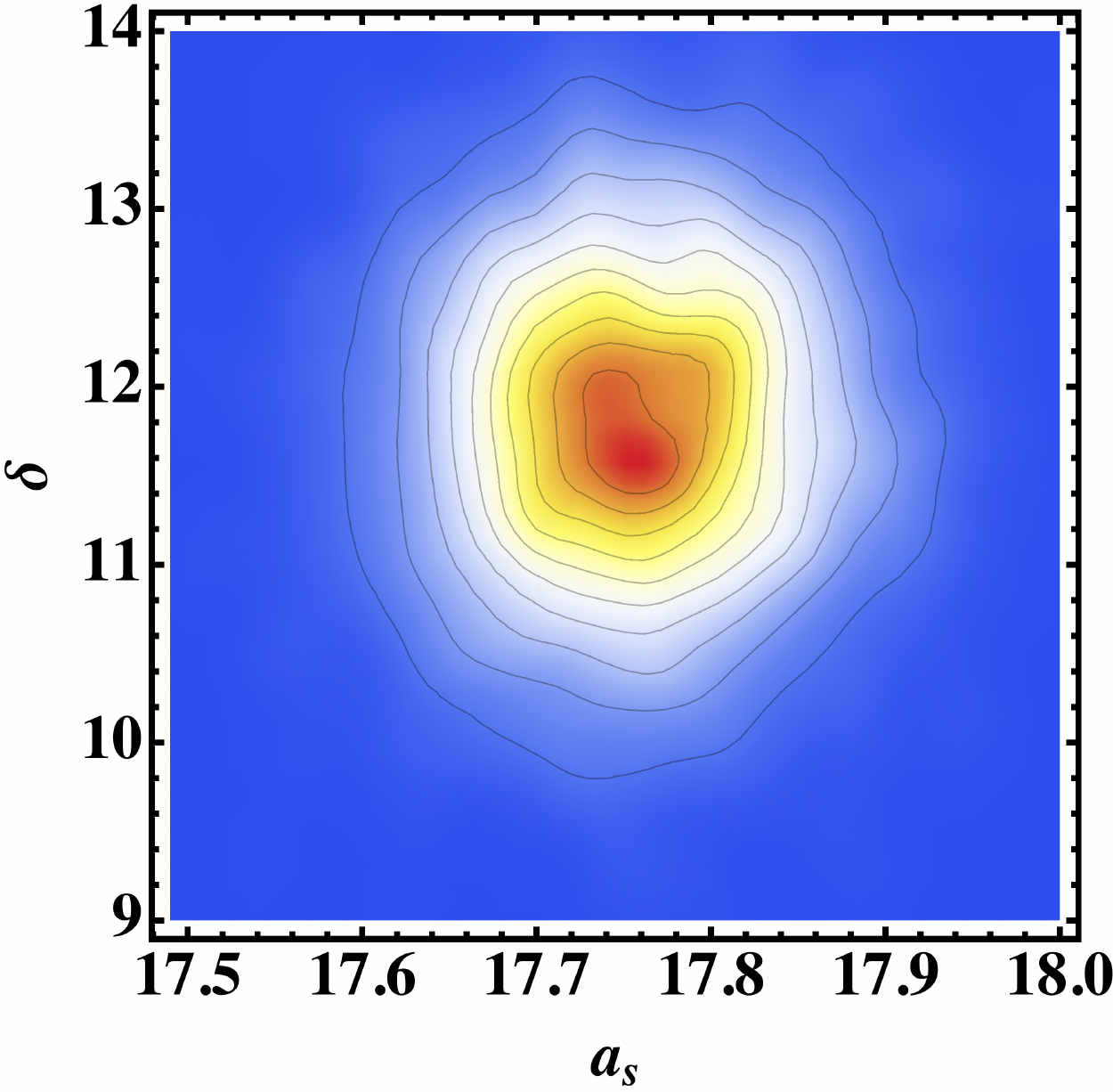}&\includegraphics[width=0.18\textwidth,angle=0]{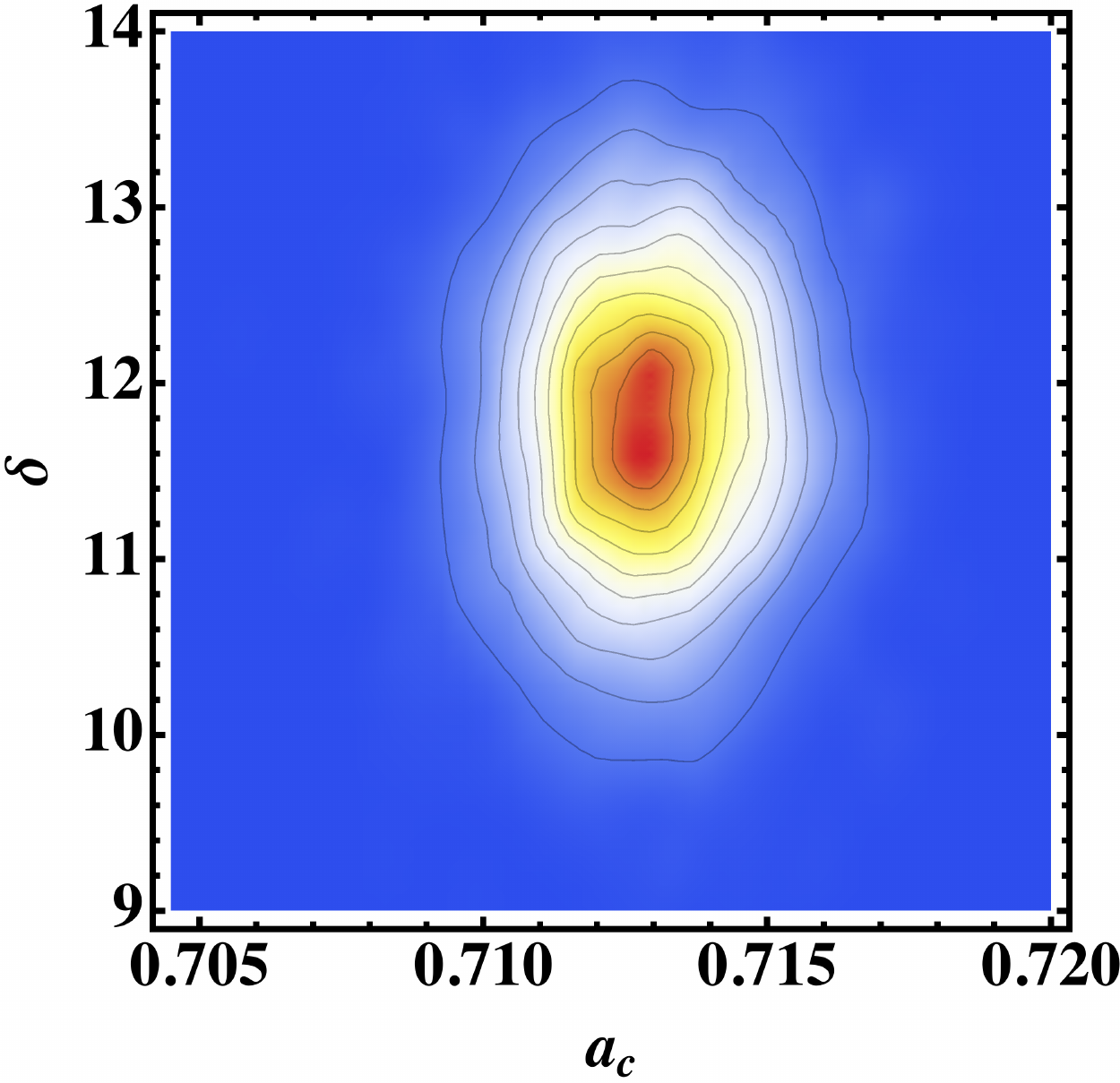}&\includegraphics[width=0.18\textwidth,angle=0]{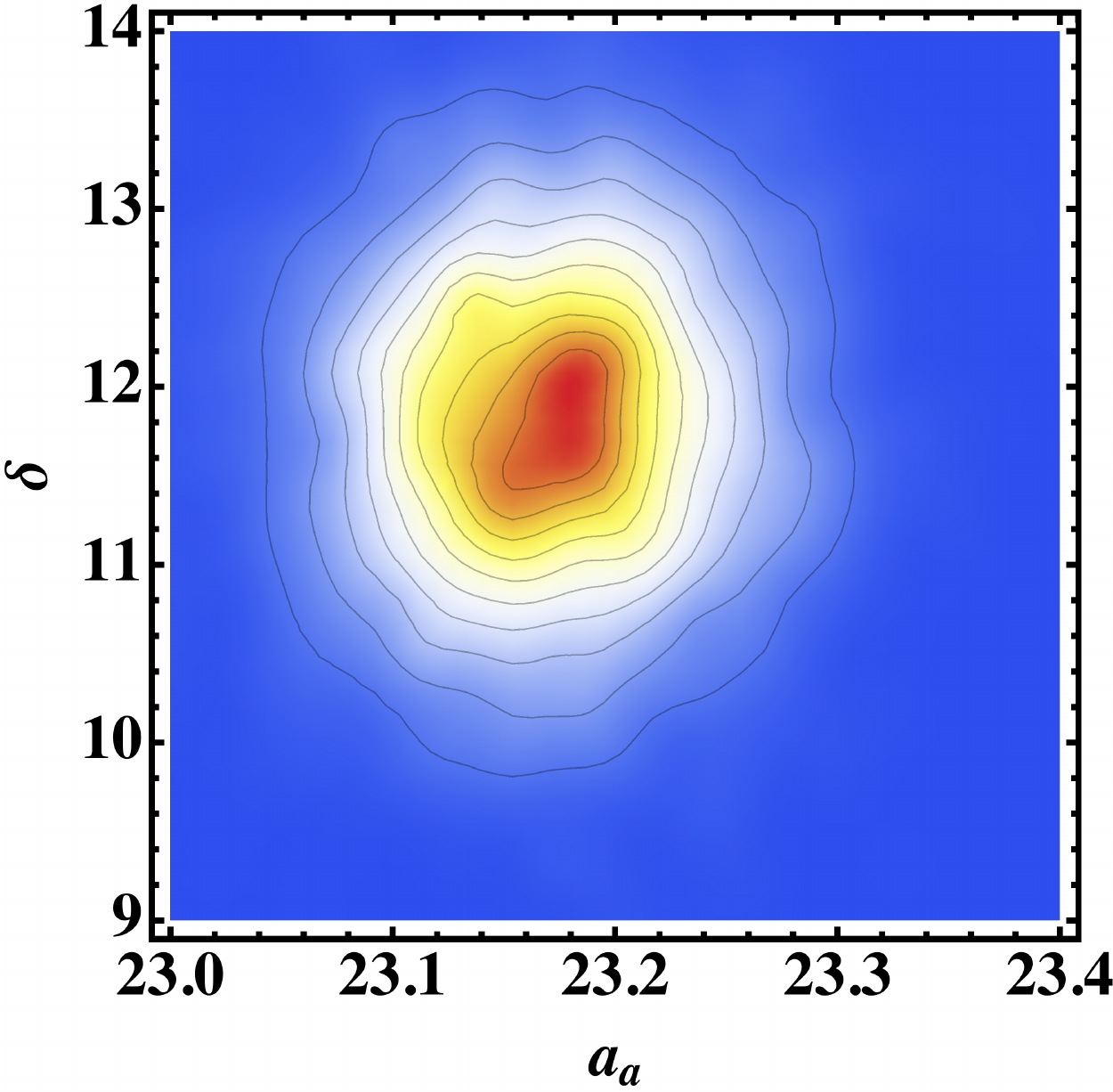}&\includegraphics[width=0.18\textwidth,angle=0]{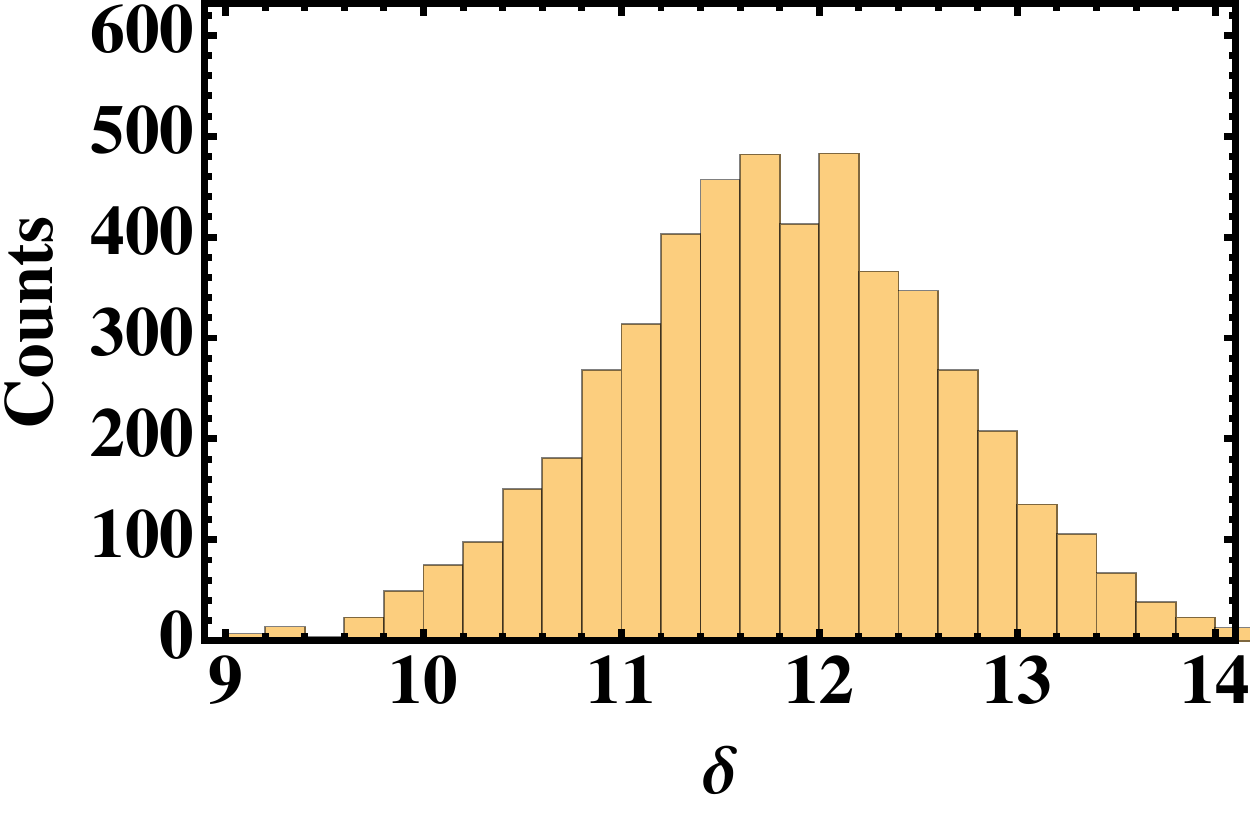}\\ 
\end{array}$
\end{center}
\caption{One and two dimensional histograms of the LD model parameters obtained with NPB assuming uncorrelated residuals. The colour code of the 2-dimensional histogram is an heat-map. The total number of counts is fixed for all the plots $N_{Boot}=5000$. See text for details.}
\label{fig:corner}
\end{figure}

I now illustrate how to apply NPB to estimate the parameters of the model in Eq.\ref{bene} The crucial aspect of NPB is to generate a new series of residuals $\mathcal{E}^*(A)$ from random sampling the original one   $\mathcal{E}(A)$ and use them to generate a new set of observables (in this case the binding energy) as 

\begin{eqnarray}
B^*_{exp}(N,Z)=B_{exp}(N,Z)+\mathcal{E}^*(N,Z)\;,
\end{eqnarray}

\noindent and then use $B_{exp}(N,Z)^*$ as a new input for a LS minimisation. This is only one possible variant of NPB, for more details I refer to Ref.~\cite{efron1994introduction}.
By performing $5000$ bootstrap samplings I derived the empirical distribution of all parameters.
I show in Fig.\ref{fig:corner} the two and one dimensional distribution of the parameters obtained with NPB. This figure is in excellent agreement with the marginalised likelihood obtained with Markov-Chain Monte Carlo in Ref.\cite{shelley2018advanced}.
From the one dimensional histogram I extract the 68\% quantile. For the first term of Eq.\ref{bene}, I obtain  $a_v=15.69_{-0.023}^{0.024}$, these values are in good agreement  with the error bars provided in Tab.\ref{tab:fit}. Similarly for the other parameters, the error bars obtained with LS and NPB are in agreement. 
This means NPB provides the same level of accuracy of a standard LS fitting as already spotted in in Ref.~\cite{bertsch2017estimating}.

The two-dimensional histograms inform us on possible correlations among parameters. Getting a cigar-shape along the diagonal of the figure means strong correlations, while a circle means low or no correlation. 
I observe that all LD parameters $a_v,a_s,a_c,a_a$ are strongly correlated among each other, while the pairing term $\delta$ is not. In this model, $\delta$ is less constrained compared to other parameters as one can also see from the typical error bars obtained in Tab.\ref{tab:fit}.
These results can be compared with the one given in Tab.\ref{tab:corr}. Using the data illustrated in Fig.\ref{fig:corner}, I have been able to extract the full covariance matrix which is in perfect agreement with the one calculated using the Hessian matrix of the $\chi^2$~\cite{bar89}. This means I did not make use of any explicit numerical derivative in parameter space.

The covariance matrix it not only useful to provide indication concerning error bars and correlations, but it can also be used to extract useful informations about the geometrical properties of the $\chi^2$ surface around the minimum. NPB by construction explores the vicinity of the minimum of Eq.\ref{chi2}, thus producing the corner plot shown in fig.\ref{fig:corner}. By analysing this figure, it is possible to assess qualitatively which directions in parameter space are more \emph{stiff}.
 I refer to Refs~\cite{gut07,nik16} for a more detailed discussion on the topic.

\subsection{Block-Bootstrap}\label{block}

As seen in Fig.\ref{resid}, ignoring the autocorrelation between residuals is quite a major approximation. In this section, I  illustrate how to correct the NPB to take into account correlations in a simple way.

As a first step, I observe that the variation in the direction N+Z is much more important than in the direction N-Z, so I define an \emph{average} residual as

\begin{eqnarray}\label{av:res}
\mathcal{E}_A(A)=\frac{1}{N_A}\sum_{Z+N=A}\mathcal{E}(N,Z)\;.
\end{eqnarray}

The result is reported in Fig.\ref{resid} as a solid line. The averaging is not washing out the signal in the data and it is used to simplify the bootstrap treatment.
The information along the variation in the N-Z is kept in  as an isotopic-dependent  variance function defined as

\begin{eqnarray}\label{varia}
\sigma^2_A=\frac{1}{N_A}\sum_{Z+N=A}\left( \mathcal{E}(N,Z)-\mathcal{E}_A(A)\right)^2\;.
\end{eqnarray}

\begin{figure}[!h]
\begin{center}
\includegraphics[width=0.38\textwidth,angle=-90]{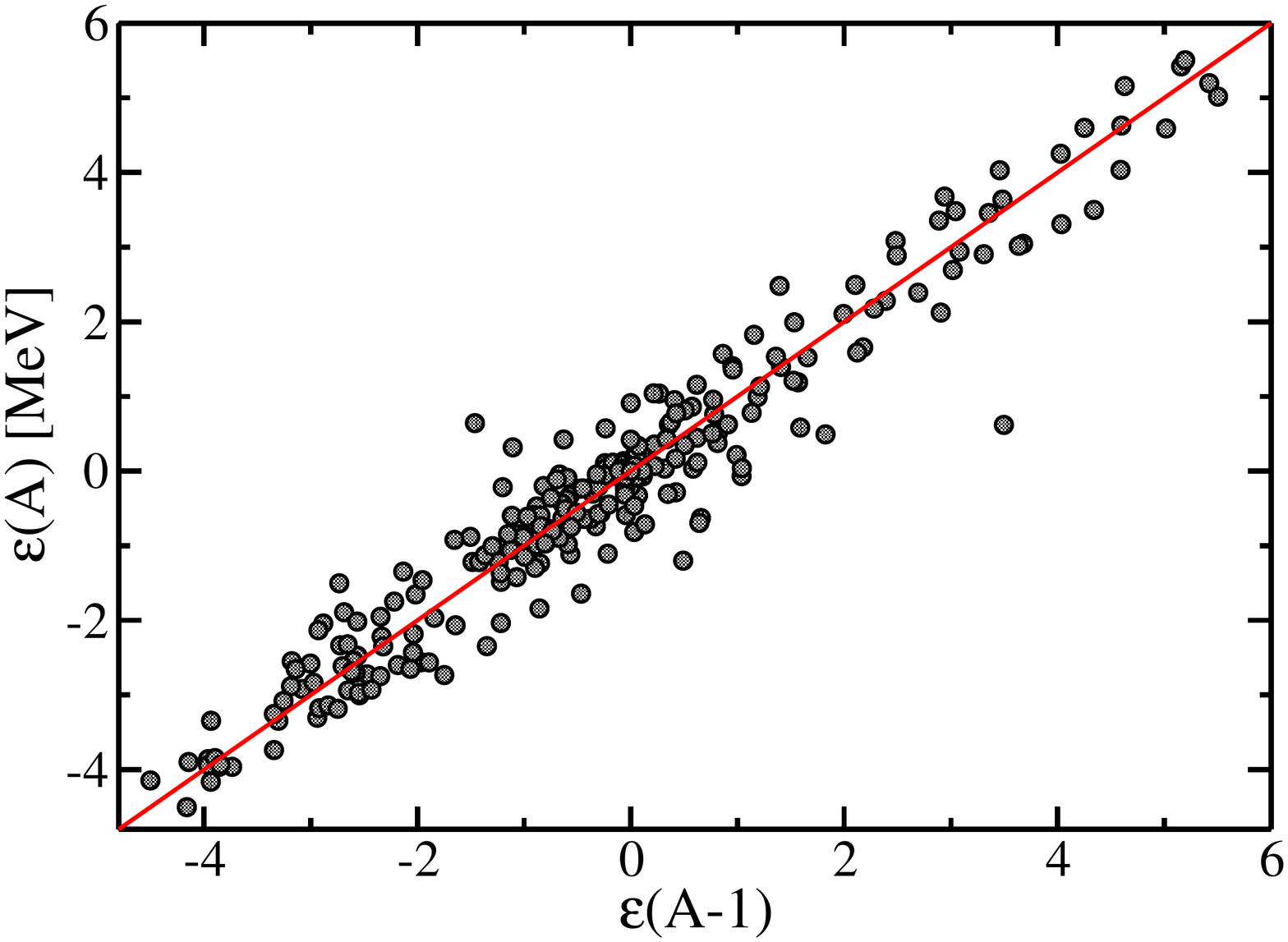}
\includegraphics[width=0.38\textwidth,angle=-90]{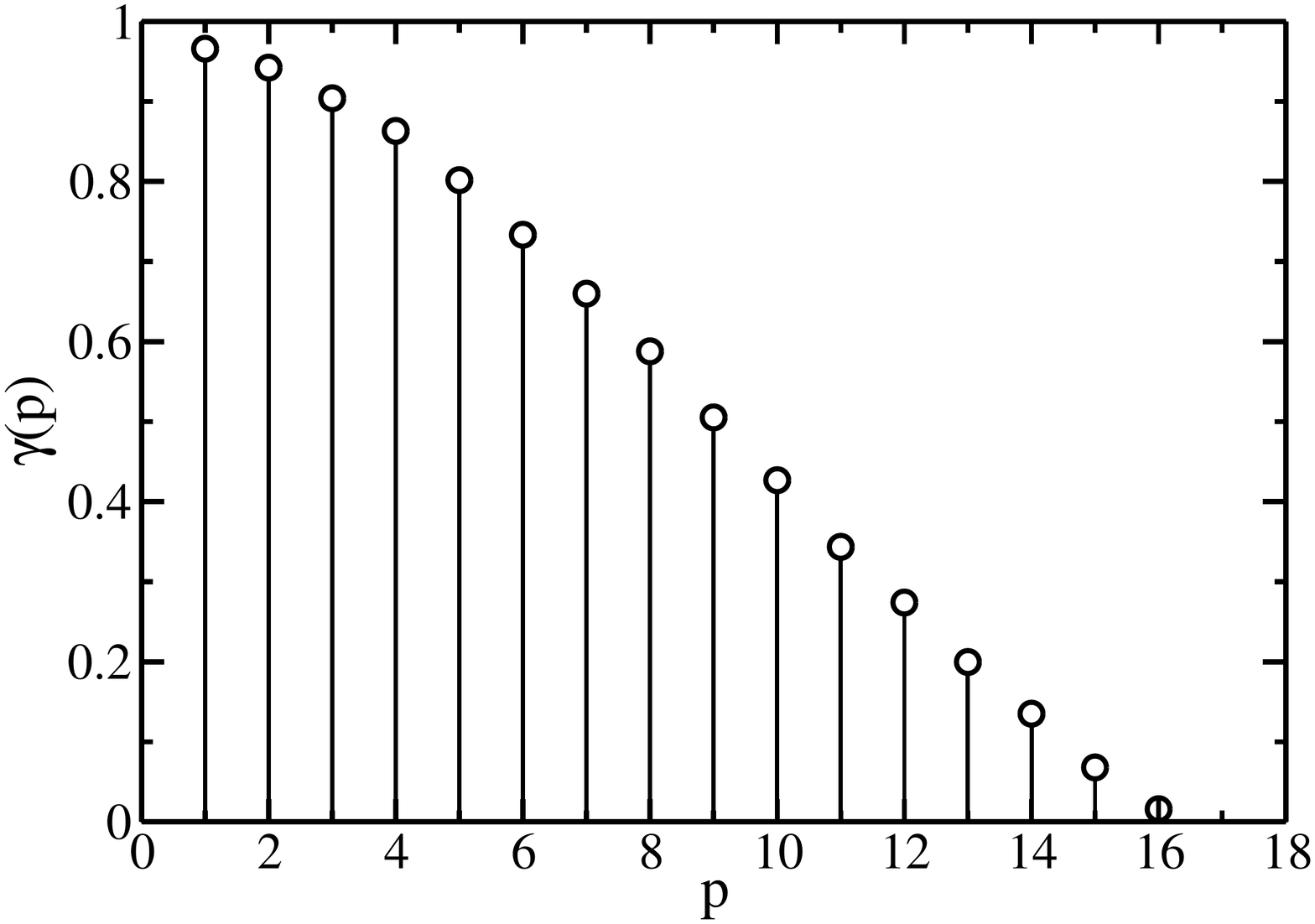}
\end{center}
\caption{(Colors online) Left panel: lag-plot of order 1 . Right panel: evolution of the self-correlation coefficient defined in Eq.\ref{selfcor} for different values of the lag.}
\label{lag}
\end{figure}

I start determining the degree of correlation between data, by drawing a lag-plot. This is a well know technique to analyse correlation in time series~\cite{brockwell2013time}. The lag-plot of lag-$p$ consists in plotting the series  against itself, but shifted by $p$ units.
In case of non-correlated residues the lag-plot should not show any pattern, \emph{i.e} one should observe a cloud of points. In Fig. \ref{lag} (left panel), I show the lag-1 plot of the averaged residuals. I observe the residuals cluster around the diagonal thus showing a strong autocorrelation. 

To quantify the correlation, I define a self-correlation coefficient as~\cite{brockwell2013time}

\begin{eqnarray}\label{selfcor}
\gamma(p)=\frac{\sum_i (X_i-\bar{X}) (Y_i-\bar{Y})}{\sigma_x\sigma_y}\;,
\end{eqnarray}

where $Y_i=X_{i-p}$ is the delayed series.
In Fig.\ref{lag} (right panel),  I present the value of such a coefficients for different lags. I observe that there is a non-negligible correlation (i.e. $\gamma(p) \ge 0.5$)  up to $p\approx8$; this means that the residual of a nucleus $A$ is strongly correlated with all other residuals within the interval $[A-8,A+8]$.
This is  in contradiction with the assumption made in Eq.\ref{chi2}, where the $\sigma^2$ matrix is assumed to be diagonal.

To take into account the presence of a signal in the residuals, one needs to introduce a minor modification to NPB. Following Ref.~\cite{kreiss2012bootstrap}, instead of resampling individual residuals, I do re-sample blocks of fixed size to preserve the correlations between data. This procedure is named Block Bootstrap (BB).

Given a data-set composed by $n$ elements $\{X_1,X_2,\dots,X_n\}$, I consider an integer $l$ satisfying $1\le l\le n$.
I define $\mathcal{B}_N$ overlapping blocks of length $l$ as

\begin{eqnarray*}
\mathcal{B}_1&=&(X_1,X_2,\dots,X_l)\\
\mathcal{B}_1&=&\phantom{X_1} (X_2,X_3,\dots,X_{l+1})\\
\dots &=&\phantom{(X_1,X_2,\dots,X_l)}\dots\\
\mathcal{B}_N&=&\phantom{(X_1,X_2,\dots,X_l),\dots}(X_{n-l+1},\dots,X_n)
\end{eqnarray*}

\noindent where $N=n-l+1$.
The uncorrelated version of bootstrap discussed in previous section is a particular case of BB when $l=1$. In this case there is no overlap between blocks and one should not use BB terminology for such a case.
I now treat the different blocks $\mathcal{B}_N$ as independent and I thus apply the standard bootstrap algorithm to them.
Once the new residuals are produced, I restore the explicit ($N,Z$) dependence by adding an additional error extracted from a Gaussian distribution with variance $\sigma^2(A)$ and defined in Eq.\ref{varia}. 
This is the same procedure followed in Ref.~\cite{bertsch2017estimating} in the case of Frequency Domain Bootstrap (FDB).

The choice of $l$ is somehow arbitrary, but in the present case case, I use the lag-plot as indication: I observe that data are correlated over an extension of at least 8 units. 
I have performed a test on the LD model using BB and various values of the length $l$ of the block.

\begin{figure}[!h]
\begin{center}
\includegraphics[width=0.38\textwidth,angle=-90]{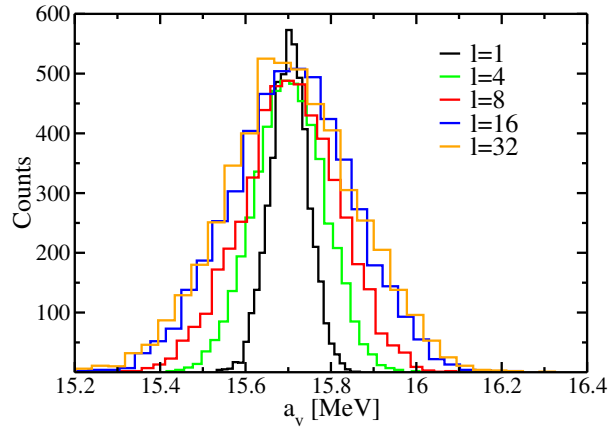}
\end{center}
\caption{(Colors online) Distribution of the $a_v$ parameter obtained using BB and different length of the block. See text for details.}
\label{bb:av}
\end{figure}

In Fig.\ref{bb:av}, I report the experimental distribution obtained using BB for the $a_v$ parameter as a function of the size of the block. 

By increasing the size of the blocks, I observe that the distributions gets wider and wider thus showing that I'm taking into account the effect of correlations. 
I see that  $l=8$ captures quite well the correlations, and beyond $l=16$ the distribution does not change anymore. This is due to the fact that the size of the blocks is getting bigger than the size of the correlations. Such a result confirms the outcomes of the lag-plot shown in Fig.\ref{lag}.

From the distributions, I now extract the new error bars of the parameters using the 68\% quantile definition.
The results are reported in Tab.\ref{tab:fitBB}. For simplicity I defined the errors to be symmetric around the mean value. This is a minor approximation that has no effect on my conclusions.

\begin{table}
\begin{center}
\begin{tabular}{c|cc}
\hline
\hline
Parameter             &  [MeV] &  Error [MeV]\\
             \hline
 $a_v$&   15.70  &       $\pm$0.14\\
  $a_s$&   17.82 &        $\pm$0.44\\
 $a_c$&  0.713 &        $\pm$0.009\\
 $a_a$   &23.20 &      $\pm$0.35\\
  $\delta$&   11.78&       $\pm$0.80 \\
\hline
\hline
\end{tabular}
\caption{Parameters  of LD models obtained using BB and $l=16$.  }
\label{tab:fitBB}
\end{center}
\end{table}

I observe that the errors are more important using this method and they are most likely more realistic than the one obtained in Tab.\ref{tab:fit}. The only exception is the error on $\delta$ parameter which remains more or less constant. This is  related to the fact that it is weakly correlated with the other terms. See Tab.\ref{tab:corr}.
The BB does not change the structure of correlations, the results of a corner plot for BB case are quite similar to the ones shown in Fig.\ref{fig:corner}.
These results are in good agreement with the one reported in Ref.~\cite{bertsch2017estimating} using FDB.
Since the covariance matrix changes due to the effect of correlation, this  affects the error propagation of the model. For example the error on the binding energy of $^{208}$Pb when I use standard bootstrap is $\pm110$ keV, while when I use BB I obtain $\pm670$ keV, thus a factor of 6 larger.
It would be now interesting to apply BB to more realistic models as the ones presented in Refs~\cite{pomorski2003nuclear,duflo1995microscopic} to study how error propagation~\cite{toi08,mcd15} of model parameters is affected.

\section{Conclusions}\label{sec:concl}

In the present article, I have discussed in detail  bootstrap methods and how they can be used to quantify uncertainties for statistical estimators.
In particular, I have shown how the minimal hypothesis behind NPB provides a better understanding of the data-set under investigation. The NPB is consistent with findings based on other methods, but it has the remarkable advantage of being able to spot possible outliers in the data-set in a very simple way, without visual investigation of the data. 
This feature may have interesting application in the case   of automatised  statistical analysis.

In Se.\ref{sec:regression}, I have shown how NPB can be used to fit parameters of a model, giving at the same time a reasonable estimation of errors and correlations without performing explicitly derivative in parameter space. 
Contrary to the standard parabolic approximation used to derive errors in a MLE method, the NPB explores the surroundings of the minimum in hyper-parameter space, thus having the possibility of finding possible defects in the hyper-surface.

To take properly into account correlations, I  have also discussed a variant of NPB named Block Bootstrap. I have proved that such a method has the same quality of results of Frequency Domain Bootstrap ~\cite{bertsch2017estimating}, but having the advantage on not using an explicit Fourier transformation of the data and thus potentially being computationally faster.
I have also highlighted that the main feature of BB and FDB is that they do not require an explicit modelling of correlations by introducing an \emph{ad-hoc} covariance matrix into the fit.

\section*{Acknowledgments}

This work has been supported by STFC Grant No. ST/P003885/1. I also acknowledge the useful discussion with P.Becker who inspired this work.

\begin{appendix}
\section{Infinite matter properties}

I report here the explicit nuclear models used in my analysis together with their basic infinite nuclear matter properties.
I refer the reader to Refs.~\cite{dut12,sellahewa2014isovector,dutra2014relativistic,davesne2016extended} for more details on the definition of infinite matter properties and how they relate to the parameters of the functionals.

\begin{table}
\begin{center}
\begin{tabular}{c|ccc||c|ccc}
\hline
\hline
  & $\rho_{0}$ [fm$^{-3}$] & $J_0$ [MeV] & $L_0$ [MeV] &  & $\rho_{0}$ [fm$^{-3}$] & $J_0$ [MeV] & $L_0$ [MeV] \\
\hline
BSK16~\cite{chamel2008further}  &  0.159 &   30.00&   34.88& SKA25S20~\cite{dut12} &  0.161 &   33.78&   63.82\\
BSK18~\cite{Chamel2010a}  &  0.159 &   30.00&   36.22 &SKB~\cite{koh76}     &  0.155 &   23.89&   47.56\\
BSK19~\cite{gor10}  &  0.160 &   30.00&   31.9&SKM~\cite{kri80}    &  0.160 &   30.75&   49.36\\
BSK20~\cite{gor10}  &  0.160 &   30.00&   37.39&SKM*~\cite{bar82}   &  0.160 &   30.03&   45.78\\
BSK21~\cite{gor10}  &  0.158 &   30.13&   46.97&SKRA~\cite{ras00}   &  0.159 &   31.32&   53.05\\
BSK22~\cite{gor13}  &  0.158 &   32.00&   68.49&SKS1~\cite{gom92}   &  0.161 &   28.75&   30.52\\
BSK23~\cite{gor13}  &  0.158 &   31.00&   57.77&SKS3~\cite{gom92}   &  0.161 &   28.84&   51.74\\
BSK24~\cite{gor13}  &  0.158 &   30.00&   46.40&SLYMR1~\cite{sad13}  &  0.155 &   33.34&   49.84\\
BSK25~\cite{gor13}  &  0.159 &   29.00&   36.90&SKSC10~\cite{onsi1994equation}  &  0.161 &   28.11&    0.21\\
BSK26 ~\cite{gor13} &  0.159 &   30.00&   37.49&SKT~\cite{ko1974microscopic}    &  0.148 &   24.90&   28.27\\
F+~\cite{Lesinski2006}     &  0.162 &   32.00&   41.54&SKT1~\cite{tondeur1984}   &  0.161 &   32.02&   56.19\\
F-  ~\cite{Lesinski2006}   &  0.162 &   32.00&   43.79&SKT1A~\cite{dut12}  &  0.161 &   32.02&   56.20\\
F0  ~\cite{Lesinski2006}   &  0.162 &   32.00&   42.41&SKT5~\cite{tondeur1984}    &  0.164 &   37.00&   98.50\\
KDE ~\cite{agr05}   &  0.164 &   31.97&   41.42&SKT5A~\cite{dut12}  &  0.164 &   37.00&   98.50\\
KDE0V ~\cite{agr05} &  0.161 &   32.98&   45.21&SKT9~\cite{tondeur1984}    &  0.160 &   29.76&   33.77\\
KDE0V1~\cite{agr05} &  0.165 &   34.58&   54.69&SKT9A~\cite{dut12}  &  0.160 &   29.76&   33.77\\
LNS1~\cite{gam11}   &  0.162 &   29.91&   30.93&SKX~\cite{brown1998new}    &  0.155 &   31.11&   33.23\\
MSK1~\cite{ton00}   &  0.157 &   30.00&   33.92&SKXCE~\cite{brown1998new}  &  0.155 &   30.21&   33.67\\
MSK2~\cite{ton00}   &  0.157 &   30.00&   33.35&SKXM~\cite{brown1998new}   &  0.159 &   31.20&   32.09\\
MSK3~\cite{ton00}   &  0.158 &   27.99&    6.83&SLY230A~\cite{cha97} &  0.160 &   31.99&   44.33\\
NRAPR~\cite{ste05}  &  0.161 &   32.78&   59.63&SLY230B~\cite{cha97} &  0.160 &   32.01&   45.97\\
RATP~\cite{ray82}   &  0.160 &   29.26&   32.43&SLY4~\cite{cha97}   &  0.160 &   32.00&   45.96\\
SEFM074~\cite{dut12} &  0.160 &   33.40&   88.71&SLY5~\cite{cha97}   &  0.161 &   32.01&   48.15\\
SEFM081~\cite{dut12} &  0.161 &   30.76&   79.38&SQMC750 &  0.171 &   33.75&   64.67\\
SEFM09~\cite{dut12} &  0.161 &   27.78&   69.95&SV~\cite{bei75}     &  0.155 &   32.82&   96.07\\
SEFM1~\cite{dut12}  &  0.161 &   24.81&   59.55&V100~\cite{pearson2001isovector}   &  0.157 &   28.00&    8.73\\
SGII~\cite{van81}   &  0.158 &   26.83&   37.66&D1M ~\cite{gor09}    &  0.164 &   28.50&   24.85\\
SGOI ~\cite{she09}  &  0.168 &   45.20&   99.76&D1S ~\cite{dec80}   &  0.163 &   31.12&   22.46\\
SI~\cite{vau72}     &  0.155 &   29.25&    1.29&DD-PC1~\cite{nik08} &  0.152 &   33.00&   70.00\\
SIII~\cite{gia80}   &  0.145 &   28.16&    9.91&DD-ME1~\cite{lal05} &  0.152 &   33.06&   55.53\\
SKA~\cite{koh76}    &  0.155 &   32.91&   74.62&DD-ME2~\cite{lal05} &  0.152 &   32.31&   51.39\\
SN2LO1~\cite{bec17} &  0.162 &   31.96&   48.89&              &                          &   & \\
\hline
\hline
\end{tabular}
\caption{Infinite matter properties of the data-set used in the current analysis.}
\label{tab:inm}
\end{center}
\end{table}

\end{appendix}

\bibliography{biblio}

\begin{thebibliography}{78}
\expandafter\ifx\csname natexlab\endcsname\relax\def\natexlab#1{#1}\fi
\expandafter\ifx\csname bibnamefont\endcsname\relax
  \def\bibnamefont#1{#1}\fi
\expandafter\ifx\csname bibfnamefont\endcsname\relax
  \def\bibfnamefont#1{#1}\fi
\expandafter\ifx\csname citenamefont\endcsname\relax
  \def\citenamefont#1{#1}\fi
\expandafter\ifx\csname url\endcsname\relax
  \def\url#1{\texttt{#1}}\fi
\expandafter\ifx\csname urlprefix\endcsname\relax\def\urlprefix{URL }\fi
\providecommand{\bibinfo}[2]{#2}
\providecommand{\eprint}[2][]{\url{#2}}

\bibitem[{\citenamefont{R.J.Barlow}(1989)}]{bar89}
\bibinfo{author}{\bibnamefont{R.J.Barlow}}, \emph{\bibinfo{title}{A Guide to
  the Use of Statistical Methods in the Physical Sciences}}
  (\bibinfo{publisher}{John Wiley}, \bibinfo{year}{1989}).

\bibitem[{\citenamefont{Dobaczewski et~al.}(2014)\citenamefont{Dobaczewski,
  Nazarewicz, and Reinhard}}]{dob14}
\bibinfo{author}{\bibfnamefont{J.}~\bibnamefont{Dobaczewski}},
  \bibinfo{author}{\bibfnamefont{W.}~\bibnamefont{Nazarewicz}},
  \bibnamefont{and} \bibinfo{author}{\bibfnamefont{P.}~\bibnamefont{Reinhard}},
  \bibinfo{journal}{Journal of Physics G: Nuclear and Particle Physics}
  \textbf{\bibinfo{volume}{41}}, \bibinfo{pages}{074001}
  (\bibinfo{year}{2014}).

\bibitem[{\citenamefont{Roca-Maza et~al.}(2015)\citenamefont{Roca-Maza, Paar,
  and Col{\`o}}}]{roca2015covariance}
\bibinfo{author}{\bibfnamefont{X.}~\bibnamefont{Roca-Maza}},
  \bibinfo{author}{\bibfnamefont{N.}~\bibnamefont{Paar}}, \bibnamefont{and}
  \bibinfo{author}{\bibfnamefont{G.}~\bibnamefont{Col{\`o}}},
  \bibinfo{journal}{Journal of Physics G: Nuclear and Particle Physics}
  \textbf{\bibinfo{volume}{42}}, \bibinfo{pages}{034033}
  (\bibinfo{year}{2015}).

\bibitem[{\citenamefont{Efron}(1979)}]{efr79}
\bibinfo{author}{\bibfnamefont{B.}~\bibnamefont{Efron}},
  \bibinfo{journal}{Annals of Statistics} \textbf{\bibinfo{volume}{7}},
  \bibinfo{pages}{1} (\bibinfo{year}{1979}).

\bibitem[{\citenamefont{Efron and Tibshirani}(1994)}]{efron1994introduction}
\bibinfo{author}{\bibfnamefont{B.}~\bibnamefont{Efron}} \bibnamefont{and}
  \bibinfo{author}{\bibfnamefont{R.~J.} \bibnamefont{Tibshirani}},
  \emph{\bibinfo{title}{An introduction to the bootstrap}}
  (\bibinfo{publisher}{CRC press}, \bibinfo{year}{1994}).

\bibitem[{\citenamefont{Davison and Hinkley}(1997)}]{dav97}
\bibinfo{author}{\bibfnamefont{A.~C.} \bibnamefont{Davison}} \bibnamefont{and}
  \bibinfo{author}{\bibfnamefont{D.~V.} \bibnamefont{Hinkley}},
  \emph{\bibinfo{title}{Bootstrap methods and their application}},
  vol.~\bibinfo{volume}{1} (\bibinfo{publisher}{Cambridge university press},
  \bibinfo{year}{1997}).

\bibitem[{\citenamefont{Chernick et~al.}(2011)\citenamefont{Chernick,
  Gonz{\'a}lez-Manteiga, Crujeiras, and Barrios}}]{chernick2011bootstrap}
\bibinfo{author}{\bibfnamefont{M.~R.} \bibnamefont{Chernick}},
  \bibinfo{author}{\bibfnamefont{W.}~\bibnamefont{Gonz{\'a}lez-Manteiga}},
  \bibinfo{author}{\bibfnamefont{R.~M.} \bibnamefont{Crujeiras}},
  \bibnamefont{and} \bibinfo{author}{\bibfnamefont{E.~B.}
  \bibnamefont{Barrios}}, in \emph{\bibinfo{booktitle}{International
  encyclopedia of statistical science}} (\bibinfo{publisher}{Springer},
  \bibinfo{year}{2011}), pp. \bibinfo{pages}{169--174}.

\bibitem[{\citenamefont{Manly}(2006)}]{manly2006randomization}
\bibinfo{author}{\bibfnamefont{B.~F.} \bibnamefont{Manly}},
  vol.~\bibinfo{volume}{70} (\bibinfo{publisher}{CRC press},
  \bibinfo{year}{2006}).

\bibitem[{\citenamefont{Chernick}(2008)}]{chernick}
\bibinfo{author}{\bibfnamefont{M.~R.} \bibnamefont{Chernick}},
  \emph{\bibinfo{title}{Bootstrap methods: a guide for practitioners and
  researchers}}, vol.~\bibinfo{volume}{1}
  (\bibinfo{publisher}{Wiley-Interscience}, \bibinfo{year}{2008}).

\bibitem[{\citenamefont{P{\'e}rez et~al.}(2014)\citenamefont{P{\'e}rez, Amaro,
  and Arriola}}]{per14}
\bibinfo{author}{\bibfnamefont{R.~N.} \bibnamefont{P{\'e}rez}},
  \bibinfo{author}{\bibfnamefont{J.}~\bibnamefont{Amaro}}, \bibnamefont{and}
  \bibinfo{author}{\bibfnamefont{E.~R.} \bibnamefont{Arriola}},
  \bibinfo{journal}{Physics Letters B} \textbf{\bibinfo{volume}{738}},
  \bibinfo{pages}{155} (\bibinfo{year}{2014}).

\bibitem[{\citenamefont{Nieves}(2000)}]{Nie00}
\bibinfo{author}{\bibfnamefont{E.}~\bibnamefont{Nieves}, \bibfnamefont{J.and
  Ruiz~Arriola}}, \bibinfo{journal}{The European Physical Journal A}
  \textbf{\bibinfo{volume}{8}}, \bibinfo{pages}{377} (\bibinfo{year}{2000}),
  ISSN \bibinfo{issn}{1434-601X}.

\bibitem[{\citenamefont{Bertsch and Bingham}(2017)}]{bertsch2017estimating}
\bibinfo{author}{\bibfnamefont{G.}~\bibnamefont{Bertsch}} \bibnamefont{and}
  \bibinfo{author}{\bibfnamefont{D.}~\bibnamefont{Bingham}},
  \bibinfo{journal}{Physical review letters} \textbf{\bibinfo{volume}{119}},
  \bibinfo{pages}{252501} (\bibinfo{year}{2017}).

\bibitem[{\citenamefont{Muir et~al.}(2018)\citenamefont{Muir, Pastore,
  Dobaczewski, and Barton}}]{muir2018bootstrap}
\bibinfo{author}{\bibfnamefont{D.}~\bibnamefont{Muir}},
  \bibinfo{author}{\bibfnamefont{A.}~\bibnamefont{Pastore}},
  \bibinfo{author}{\bibfnamefont{J.}~\bibnamefont{Dobaczewski}},
  \bibnamefont{and} \bibinfo{author}{\bibfnamefont{C.}~\bibnamefont{Barton}},
  \bibinfo{journal}{Acta Physica Polonica B} \textbf{\bibinfo{volume}{49}}
  (\bibinfo{year}{2018}).

\bibitem[{\citenamefont{Pasquini et~al.}(2018)\citenamefont{Pasquini, Pedroni,
  and Sconfietti}}]{pasq18}
\bibinfo{author}{\bibfnamefont{B.}~\bibnamefont{Pasquini}},
  \bibinfo{author}{\bibfnamefont{P.}~\bibnamefont{Pedroni}}, \bibnamefont{and}
  \bibinfo{author}{\bibfnamefont{S.}~\bibnamefont{Sconfietti}},
  \bibinfo{journal}{Physical Review C} \textbf{\bibinfo{volume}{98}},
  \bibinfo{pages}{015204} (\bibinfo{year}{2018}).

\bibitem[{\citenamefont{Fisher and Hall}(1990)}]{fisher1990new}
\bibinfo{author}{\bibfnamefont{N.~I.} \bibnamefont{Fisher}} \bibnamefont{and}
  \bibinfo{author}{\bibfnamefont{P.}~\bibnamefont{Hall}},
  \bibinfo{journal}{Geophysical Journal International}
  \textbf{\bibinfo{volume}{101}}, \bibinfo{pages}{305} (\bibinfo{year}{1990}).

\bibitem[{\citenamefont{Miller}(1974)}]{mil74}
\bibinfo{author}{\bibfnamefont{R.~G.} \bibnamefont{Miller}},
  \bibinfo{journal}{Biometrika} \textbf{\bibinfo{volume}{61}},
  \bibinfo{pages}{1} (\bibinfo{year}{1974}).

\bibitem[{\citenamefont{Tukey}(1958)}]{tuk58}
\bibinfo{author}{\bibfnamefont{J.}~\bibnamefont{Tukey}}, \bibinfo{journal}{Ann.
  Math. Statist.} \textbf{\bibinfo{volume}{29}}, \bibinfo{pages}{614}
  (\bibinfo{year}{1958}).

\bibitem[{\citenamefont{Corey et~al.}(1998)\citenamefont{Corey, Dunlap, and
  Burke}}]{cor98}
\bibinfo{author}{\bibfnamefont{D.~M.} \bibnamefont{Corey}},
  \bibinfo{author}{\bibfnamefont{W.~P.} \bibnamefont{Dunlap}},
  \bibnamefont{and} \bibinfo{author}{\bibfnamefont{M.~J.} \bibnamefont{Burke}},
  \bibinfo{journal}{The Journal of general psychology}
  \textbf{\bibinfo{volume}{125}}, \bibinfo{pages}{245} (\bibinfo{year}{1998}).

\bibitem[{\citenamefont{Centelles et~al.}(2009)\citenamefont{Centelles,
  Roca-Maza, Vi\~nas, and Warda}}]{cen09}
\bibinfo{author}{\bibfnamefont{M.}~\bibnamefont{Centelles}},
  \bibinfo{author}{\bibfnamefont{X.}~\bibnamefont{Roca-Maza}},
  \bibinfo{author}{\bibfnamefont{X.}~\bibnamefont{Vi\~nas}}, \bibnamefont{and}
  \bibinfo{author}{\bibfnamefont{M.}~\bibnamefont{Warda}},
  \bibinfo{journal}{Phys. Rev. Lett.} \textbf{\bibinfo{volume}{102}},
  \bibinfo{pages}{122502} (\bibinfo{year}{2009}),
  \urlprefix\url{http://link.aps.org/doi/10.1103/PhysRevLett.102.122502}.

\bibitem[{\citenamefont{Roca-Maza et~al.}(2011)\citenamefont{Roca-Maza,
  Centelles, Vi\~nas, and Warda}}]{roc11}
\bibinfo{author}{\bibfnamefont{X.}~\bibnamefont{Roca-Maza}},
  \bibinfo{author}{\bibfnamefont{M.}~\bibnamefont{Centelles}},
  \bibinfo{author}{\bibfnamefont{X.}~\bibnamefont{Vi\~nas}}, \bibnamefont{and}
  \bibinfo{author}{\bibfnamefont{M.}~\bibnamefont{Warda}},
  \bibinfo{journal}{Phys. Rev. Lett.} \textbf{\bibinfo{volume}{106}},
  \bibinfo{pages}{252501} (\bibinfo{year}{2011}),
  \urlprefix\url{http://link.aps.org/doi/10.1103/PhysRevLett.106.252501}.

\bibitem[{\citenamefont{Perli{\'n}ska et~al.}(2004)\citenamefont{Perli{\'n}ska,
  Rohozi{\'n}ski, Dobaczewski, and Nazarewicz}}]{per04}
\bibinfo{author}{\bibfnamefont{E.}~\bibnamefont{Perli{\'n}ska}},
  \bibinfo{author}{\bibfnamefont{S.}~\bibnamefont{Rohozi{\'n}ski}},
  \bibinfo{author}{\bibfnamefont{J.}~\bibnamefont{Dobaczewski}},
  \bibnamefont{and}
  \bibinfo{author}{\bibfnamefont{W.}~\bibnamefont{Nazarewicz}},
  \bibinfo{journal}{Physical Review C} \textbf{\bibinfo{volume}{69}},
  \bibinfo{pages}{014316} (\bibinfo{year}{2004}).

\bibitem[{\citenamefont{Decharg{\'e} and Gogny}(1980)}]{dec80}
\bibinfo{author}{\bibfnamefont{J.}~\bibnamefont{Decharg{\'e}}}
  \bibnamefont{and} \bibinfo{author}{\bibfnamefont{D.}~\bibnamefont{Gogny}},
  \bibinfo{journal}{Physical Review C} \textbf{\bibinfo{volume}{21}},
  \bibinfo{pages}{1568} (\bibinfo{year}{1980}).

\bibitem[{\citenamefont{Bender et~al.}(2003)\citenamefont{Bender, Heenen, and
  Reinhard}}]{ben03}
\bibinfo{author}{\bibfnamefont{M.}~\bibnamefont{Bender}},
  \bibinfo{author}{\bibfnamefont{P.-H.} \bibnamefont{Heenen}},
  \bibnamefont{and} \bibinfo{author}{\bibfnamefont{P.-G.}
  \bibnamefont{Reinhard}}, \bibinfo{journal}{Reviews of Modern Physics}
  \textbf{\bibinfo{volume}{75}}, \bibinfo{pages}{121} (\bibinfo{year}{2003}).

\bibitem[{\citenamefont{Nik{\v{s}}i{\'c}
  et~al.}(2008)\citenamefont{Nik{\v{s}}i{\'c}, Vretenar, and Ring}}]{nik08}
\bibinfo{author}{\bibfnamefont{T.}~\bibnamefont{Nik{\v{s}}i{\'c}}},
  \bibinfo{author}{\bibfnamefont{D.}~\bibnamefont{Vretenar}}, \bibnamefont{and}
  \bibinfo{author}{\bibfnamefont{P.}~\bibnamefont{Ring}},
  \bibinfo{journal}{Physical Review C} \textbf{\bibinfo{volume}{78}},
  \bibinfo{pages}{034318} (\bibinfo{year}{2008}).

\bibitem[{\citenamefont{Warda et~al.}(2010)\citenamefont{Warda, Vi\~nas,
  Roca-Maza, and Centelles}}]{war10}
\bibinfo{author}{\bibfnamefont{M.}~\bibnamefont{Warda}},
  \bibinfo{author}{\bibfnamefont{X.}~\bibnamefont{Vi\~nas}},
  \bibinfo{author}{\bibfnamefont{X.}~\bibnamefont{Roca-Maza}},
  \bibnamefont{and}
  \bibinfo{author}{\bibfnamefont{M.}~\bibnamefont{Centelles}},
  \bibinfo{journal}{Phys. Rev. C} \textbf{\bibinfo{volume}{81}},
  \bibinfo{pages}{054309} (\bibinfo{year}{2010}),
  \urlprefix\url{http://link.aps.org/doi/10.1103/PhysRevC.81.054309}.

\bibitem[{\citenamefont{Tarbert et~al.}(2014)\citenamefont{Tarbert, Watts,
  Glazier, Aguar, Ahrens, Annand, Arends, Beck, Bekrenev, Boillat
  et~al.}}]{tar14}
\bibinfo{author}{\bibfnamefont{C.}~\bibnamefont{Tarbert}},
  \bibinfo{author}{\bibfnamefont{D.}~\bibnamefont{Watts}},
  \bibinfo{author}{\bibfnamefont{D.}~\bibnamefont{Glazier}},
  \bibinfo{author}{\bibfnamefont{P.}~\bibnamefont{Aguar}},
  \bibinfo{author}{\bibfnamefont{J.}~\bibnamefont{Ahrens}},
  \bibinfo{author}{\bibfnamefont{J.}~\bibnamefont{Annand}},
  \bibinfo{author}{\bibfnamefont{H.}~\bibnamefont{Arends}},
  \bibinfo{author}{\bibfnamefont{R.}~\bibnamefont{Beck}},
  \bibinfo{author}{\bibfnamefont{V.}~\bibnamefont{Bekrenev}},
  \bibinfo{author}{\bibfnamefont{B.}~\bibnamefont{Boillat}},
  \bibnamefont{et~al.}, \bibinfo{journal}{Physical review letters}
  \textbf{\bibinfo{volume}{112}}, \bibinfo{pages}{242502}
  (\bibinfo{year}{2014}).

\bibitem[{\citenamefont{Dutra et~al.}(2012)\citenamefont{Dutra,
  Louren\ifmmode~\mbox{\c{c}}\else \c{c}\fi{}o, S\'a~Martins, Delfino, Stone,
  and Stevenson}}]{dut12}
\bibinfo{author}{\bibfnamefont{M.}~\bibnamefont{Dutra}},
  \bibinfo{author}{\bibfnamefont{O.}~\bibnamefont{Louren\ifmmode~\mbox{\c{c}}\else
  \c{c}\fi{}o}}, \bibinfo{author}{\bibfnamefont{J.~S.}
  \bibnamefont{S\'a~Martins}},
  \bibinfo{author}{\bibfnamefont{A.}~\bibnamefont{Delfino}},
  \bibinfo{author}{\bibfnamefont{J.~R.} \bibnamefont{Stone}}, \bibnamefont{and}
  \bibinfo{author}{\bibfnamefont{P.~D.} \bibnamefont{Stevenson}},
  \bibinfo{journal}{Phys. Rev. C} \textbf{\bibinfo{volume}{85}},
  \bibinfo{pages}{035201} (\bibinfo{year}{2012}),
  \urlprefix\url{http://link.aps.org/doi/10.1103/PhysRevC.85.035201}.

\bibitem[{\citenamefont{Shen et~al.}(2009)\citenamefont{Shen, Han, Guo
  et~al.}}]{she09}
\bibinfo{author}{\bibfnamefont{Q.-b.} \bibnamefont{Shen}},
  \bibinfo{author}{\bibfnamefont{Y.-l.} \bibnamefont{Han}},
  \bibinfo{author}{\bibfnamefont{H.-r.} \bibnamefont{Guo}},
  \bibnamefont{et~al.}, \bibinfo{journal}{Physical Review C}
  \textbf{\bibinfo{volume}{80}}, \bibinfo{pages}{024604}
  (\bibinfo{year}{2009}).

\bibitem[{\citenamefont{Reinhard and Flocard}(1995)}]{reinhard1995nuclear}
\bibinfo{author}{\bibfnamefont{P.-G.} \bibnamefont{Reinhard}} \bibnamefont{and}
  \bibinfo{author}{\bibfnamefont{H.}~\bibnamefont{Flocard}},
  \bibinfo{journal}{Nuclear Physics A} \textbf{\bibinfo{volume}{584}},
  \bibinfo{pages}{467} (\bibinfo{year}{1995}).

\bibitem[{\citenamefont{Rubin}(1981)}]{rubin1981bayesian}
\bibinfo{author}{\bibfnamefont{D.~B.} \bibnamefont{Rubin}},
  \bibinfo{journal}{The annals of statistics} pp. \bibinfo{pages}{130--134}
  (\bibinfo{year}{1981}).

\bibitem[{\citenamefont{Boos and Monahan}(1986)}]{boos1986bootstrap}
\bibinfo{author}{\bibfnamefont{D.~D.} \bibnamefont{Boos}} \bibnamefont{and}
  \bibinfo{author}{\bibfnamefont{J.~F.} \bibnamefont{Monahan}},
  \bibinfo{journal}{Biometrika} \textbf{\bibinfo{volume}{73}},
  \bibinfo{pages}{77} (\bibinfo{year}{1986}).

\bibitem[{\citenamefont{Efron and Stein}(1981)}]{efron1981jackknife}
\bibinfo{author}{\bibfnamefont{B.}~\bibnamefont{Efron}} \bibnamefont{and}
  \bibinfo{author}{\bibfnamefont{C.}~\bibnamefont{Stein}},
  \bibinfo{journal}{The Annals of Statistics} pp. \bibinfo{pages}{586--596}
  (\bibinfo{year}{1981}).

\bibitem[{\citenamefont{Jodon}(2014)}]{robin}
\bibinfo{author}{\bibfnamefont{R.}~\bibnamefont{Jodon}}, Ph.D. thesis,
  \bibinfo{school}{Universit{\'e} de Lyon} (\bibinfo{year}{2014}).

\bibitem[{\citenamefont{Krane and Halliday}(1988)}]{krane1988introductory}
\bibinfo{author}{\bibfnamefont{K.~S.} \bibnamefont{Krane}} \bibnamefont{and}
  \bibinfo{author}{\bibfnamefont{D.}~\bibnamefont{Halliday}},
  \emph{\bibinfo{title}{Introductory nuclear physics}}, vol.
  \bibinfo{volume}{465} (\bibinfo{publisher}{Wiley New York},
  \bibinfo{year}{1988}).

\bibitem[{\citenamefont{Wang et~al.}(2012)\citenamefont{Wang, Audi, Wapstra,
  Kondev, MacCormick, Xu, and Pfeiffer}}]{wang2012ame2012}
\bibinfo{author}{\bibfnamefont{M.}~\bibnamefont{Wang}},
  \bibinfo{author}{\bibfnamefont{G.}~\bibnamefont{Audi}},
  \bibinfo{author}{\bibfnamefont{A.}~\bibnamefont{Wapstra}},
  \bibinfo{author}{\bibfnamefont{F.}~\bibnamefont{Kondev}},
  \bibinfo{author}{\bibfnamefont{M.}~\bibnamefont{MacCormick}},
  \bibinfo{author}{\bibfnamefont{X.}~\bibnamefont{Xu}}, \bibnamefont{and}
  \bibinfo{author}{\bibfnamefont{B.}~\bibnamefont{Pfeiffer}},
  \bibinfo{journal}{Chinese Physics C} \textbf{\bibinfo{volume}{36}},
  \bibinfo{pages}{1603} (\bibinfo{year}{2012}).

\bibitem[{\citenamefont{P{\'e}rez et~al.}(2015)\citenamefont{P{\'e}rez, Amaro,
  and Arriola}}]{perez2015error}
\bibinfo{author}{\bibfnamefont{R.~N.} \bibnamefont{P{\'e}rez}},
  \bibinfo{author}{\bibfnamefont{J.}~\bibnamefont{Amaro}}, \bibnamefont{and}
  \bibinfo{author}{\bibfnamefont{E.~R.} \bibnamefont{Arriola}},
  \bibinfo{journal}{Journal of Physics G: Nuclear and Particle Physics}
  \textbf{\bibinfo{volume}{42}}, \bibinfo{pages}{034013}
  (\bibinfo{year}{2015}).

\bibitem[{\citenamefont{Kl{\"u}pfel et~al.}(2009)\citenamefont{Kl{\"u}pfel,
  Reinhard, B{\"u}rvenich, and Maruhn}}]{klupfel2009variations}
\bibinfo{author}{\bibfnamefont{P.}~\bibnamefont{Kl{\"u}pfel}},
  \bibinfo{author}{\bibfnamefont{P.-G.} \bibnamefont{Reinhard}},
  \bibinfo{author}{\bibfnamefont{T.}~\bibnamefont{B{\"u}rvenich}},
  \bibnamefont{and} \bibinfo{author}{\bibfnamefont{J.}~\bibnamefont{Maruhn}},
  \bibinfo{journal}{Physical Review C} \textbf{\bibinfo{volume}{79}},
  \bibinfo{pages}{034310} (\bibinfo{year}{2009}).

\bibitem[{\citenamefont{Pomorski and Dudek}(2003)}]{pomorski2003nuclear}
\bibinfo{author}{\bibfnamefont{K.}~\bibnamefont{Pomorski}} \bibnamefont{and}
  \bibinfo{author}{\bibfnamefont{J.}~\bibnamefont{Dudek}},
  \bibinfo{journal}{Physical Review C} \textbf{\bibinfo{volume}{67}},
  \bibinfo{pages}{044316} (\bibinfo{year}{2003}).

\bibitem[{\citenamefont{Shelley et~al.}(2018)\citenamefont{Shelley, Becker,
  Gration, and Pastore}}]{shelley2018advanced}
\bibinfo{author}{\bibfnamefont{M.}~\bibnamefont{Shelley}},
  \bibinfo{author}{\bibfnamefont{P.}~\bibnamefont{Becker}},
  \bibinfo{author}{\bibfnamefont{A.}~\bibnamefont{Gration}}, \bibnamefont{and}
  \bibinfo{author}{\bibfnamefont{A.}~\bibnamefont{Pastore}},
  \bibinfo{journal}{arXiv preprint arXiv:1811.09130}  (\bibinfo{year}{2018}).

\bibitem[{\citenamefont{Gutenkunst et~al.}(2007)\citenamefont{Gutenkunst,
  Waterfall, Casey, Brown, Myers, and Sethna}}]{gut07}
\bibinfo{author}{\bibfnamefont{R.~N.} \bibnamefont{Gutenkunst}},
  \bibinfo{author}{\bibfnamefont{J.~J.} \bibnamefont{Waterfall}},
  \bibinfo{author}{\bibfnamefont{F.~P.} \bibnamefont{Casey}},
  \bibinfo{author}{\bibfnamefont{K.~S.} \bibnamefont{Brown}},
  \bibinfo{author}{\bibfnamefont{C.~R.} \bibnamefont{Myers}}, \bibnamefont{and}
  \bibinfo{author}{\bibfnamefont{J.~P.} \bibnamefont{Sethna}},
  \bibinfo{journal}{PLoS computational biology} \textbf{\bibinfo{volume}{3}},
  \bibinfo{pages}{e189} (\bibinfo{year}{2007}).

\bibitem[{\citenamefont{Nik{\v{s}}i{\'c} and Vretenar}(2016)}]{nik16}
\bibinfo{author}{\bibfnamefont{T.}~\bibnamefont{Nik{\v{s}}i{\'c}}}
  \bibnamefont{and} \bibinfo{author}{\bibfnamefont{D.}~\bibnamefont{Vretenar}},
  \bibinfo{journal}{Physical Review C} \textbf{\bibinfo{volume}{94}},
  \bibinfo{pages}{024333} (\bibinfo{year}{2016}).

\bibitem[{\citenamefont{Brockwell and Davis}(2013)}]{brockwell2013time}
\bibinfo{author}{\bibfnamefont{P.~J.} \bibnamefont{Brockwell}}
  \bibnamefont{and} \bibinfo{author}{\bibfnamefont{R.~A.} \bibnamefont{Davis}},
  \emph{\bibinfo{title}{Time series: theory and methods}}
  (\bibinfo{publisher}{Springer Science \& Business Media},
  \bibinfo{year}{2013}).

\bibitem[{\citenamefont{Kreiss and Lahiri}(2012)}]{kreiss2012bootstrap}
\bibinfo{author}{\bibfnamefont{J.-P.} \bibnamefont{Kreiss}} \bibnamefont{and}
  \bibinfo{author}{\bibfnamefont{S.~N.} \bibnamefont{Lahiri}}, in
  \emph{\bibinfo{booktitle}{Handbook of statistics}}
  (\bibinfo{publisher}{Elsevier}, \bibinfo{year}{2012}),
  vol.~\bibinfo{volume}{30}, pp. \bibinfo{pages}{3--26}.

\bibitem[{\citenamefont{Duflo and Zuker}(1995)}]{duflo1995microscopic}
\bibinfo{author}{\bibfnamefont{J.}~\bibnamefont{Duflo}} \bibnamefont{and}
  \bibinfo{author}{\bibfnamefont{A.}~\bibnamefont{Zuker}},
  \bibinfo{journal}{Physical Review C} \textbf{\bibinfo{volume}{52}},
  \bibinfo{pages}{R23} (\bibinfo{year}{1995}).

\bibitem[{\citenamefont{Toivanen et~al.}(2008)\citenamefont{Toivanen,
  Dobaczewski, Kortelainen, and Mizuyama}}]{toi08}
\bibinfo{author}{\bibfnamefont{J.}~\bibnamefont{Toivanen}},
  \bibinfo{author}{\bibfnamefont{J.}~\bibnamefont{Dobaczewski}},
  \bibinfo{author}{\bibfnamefont{M.}~\bibnamefont{Kortelainen}},
  \bibnamefont{and} \bibinfo{author}{\bibfnamefont{K.}~\bibnamefont{Mizuyama}},
  \bibinfo{journal}{Physical Review C} \textbf{\bibinfo{volume}{78}},
  \bibinfo{pages}{034306} (\bibinfo{year}{2008}).

\bibitem[{\citenamefont{McDonnell et~al.}(2015)\citenamefont{McDonnell,
  Schunck, Higdon, Sarich, Wild, and Nazarewicz}}]{mcd15}
\bibinfo{author}{\bibfnamefont{J.}~\bibnamefont{McDonnell}},
  \bibinfo{author}{\bibfnamefont{N.}~\bibnamefont{Schunck}},
  \bibinfo{author}{\bibfnamefont{D.}~\bibnamefont{Higdon}},
  \bibinfo{author}{\bibfnamefont{J.}~\bibnamefont{Sarich}},
  \bibinfo{author}{\bibfnamefont{S.}~\bibnamefont{Wild}}, \bibnamefont{and}
  \bibinfo{author}{\bibfnamefont{W.}~\bibnamefont{Nazarewicz}},
  \bibinfo{journal}{Physical review letters} \textbf{\bibinfo{volume}{114}},
  \bibinfo{pages}{122501} (\bibinfo{year}{2015}).

\bibitem[{\citenamefont{Sellahewa and Rios}(2014)}]{sellahewa2014isovector}
\bibinfo{author}{\bibfnamefont{R.}~\bibnamefont{Sellahewa}} \bibnamefont{and}
  \bibinfo{author}{\bibfnamefont{A.}~\bibnamefont{Rios}},
  \bibinfo{journal}{Physical Review C} \textbf{\bibinfo{volume}{90}},
  \bibinfo{pages}{054327} (\bibinfo{year}{2014}).

\bibitem[{\citenamefont{Dutra et~al.}(2014)\citenamefont{Dutra, Louren{\c{c}}o,
  Avancini, Carlson, Delfino, Menezes, Provid{\^e}ncia, Typel, and
  Stone}}]{dutra2014relativistic}
\bibinfo{author}{\bibfnamefont{M.}~\bibnamefont{Dutra}},
  \bibinfo{author}{\bibfnamefont{O.}~\bibnamefont{Louren{\c{c}}o}},
  \bibinfo{author}{\bibfnamefont{S.}~\bibnamefont{Avancini}},
  \bibinfo{author}{\bibfnamefont{B.}~\bibnamefont{Carlson}},
  \bibinfo{author}{\bibfnamefont{A.}~\bibnamefont{Delfino}},
  \bibinfo{author}{\bibfnamefont{D.}~\bibnamefont{Menezes}},
  \bibinfo{author}{\bibfnamefont{C.}~\bibnamefont{Provid{\^e}ncia}},
  \bibinfo{author}{\bibfnamefont{S.}~\bibnamefont{Typel}}, \bibnamefont{and}
  \bibinfo{author}{\bibfnamefont{J.}~\bibnamefont{Stone}},
  \bibinfo{journal}{Physical Review C} \textbf{\bibinfo{volume}{90}},
  \bibinfo{pages}{055203} (\bibinfo{year}{2014}).

\bibitem[{\citenamefont{Davesne et~al.}(2016)\citenamefont{Davesne, Pastore,
  and Navarro}}]{davesne2016extended}
\bibinfo{author}{\bibfnamefont{D.}~\bibnamefont{Davesne}},
  \bibinfo{author}{\bibfnamefont{A.}~\bibnamefont{Pastore}}, \bibnamefont{and}
  \bibinfo{author}{\bibfnamefont{J.}~\bibnamefont{Navarro}},
  \bibinfo{journal}{Astronomy \& Astrophysics} \textbf{\bibinfo{volume}{585}},
  \bibinfo{pages}{A83} (\bibinfo{year}{2016}).

\bibitem[{\citenamefont{Chamel et~al.}(2008)\citenamefont{Chamel, Goriely, and
  Pearson}}]{chamel2008further}
\bibinfo{author}{\bibfnamefont{N.}~\bibnamefont{Chamel}},
  \bibinfo{author}{\bibfnamefont{S.}~\bibnamefont{Goriely}}, \bibnamefont{and}
  \bibinfo{author}{\bibfnamefont{J.}~\bibnamefont{Pearson}},
  \bibinfo{journal}{Nuclear Physics A} \textbf{\bibinfo{volume}{812}},
  \bibinfo{pages}{72} (\bibinfo{year}{2008}).

\bibitem[{\citenamefont{Chamel and Goriely}(2010)}]{Chamel2010a}
\bibinfo{author}{\bibfnamefont{N.}~\bibnamefont{Chamel}} \bibnamefont{and}
  \bibinfo{author}{\bibfnamefont{S.}~\bibnamefont{Goriely}},
  \bibinfo{journal}{Physical Review C} \textbf{\bibinfo{volume}{82}},
  \bibinfo{pages}{045804} (\bibinfo{year}{2010}), ISSN
  \bibinfo{issn}{0556-2813}.

\bibitem[{\citenamefont{K{\"o}hler}(1976)}]{koh76}
\bibinfo{author}{\bibfnamefont{H.}~\bibnamefont{K{\"o}hler}},
  \bibinfo{journal}{Nuclear Physics A} \textbf{\bibinfo{volume}{258}},
  \bibinfo{pages}{301} (\bibinfo{year}{1976}).

\bibitem[{\citenamefont{Goriely et~al.}(2010)\citenamefont{Goriely, Chamel, and
  Pearson}}]{gor10}
\bibinfo{author}{\bibfnamefont{S.}~\bibnamefont{Goriely}},
  \bibinfo{author}{\bibfnamefont{N.}~\bibnamefont{Chamel}}, \bibnamefont{and}
  \bibinfo{author}{\bibfnamefont{J.~M.} \bibnamefont{Pearson}},
  \bibinfo{journal}{Phys. Rev. C} \textbf{\bibinfo{volume}{82}},
  \bibinfo{pages}{035804} (\bibinfo{year}{2010}),
  \urlprefix\url{https://link.aps.org/doi/10.1103/PhysRevC.82.035804}.

\bibitem[{\citenamefont{Krivine et~al.}(1980)\citenamefont{Krivine, Treiner,
  and Bohigas}}]{kri80}
\bibinfo{author}{\bibfnamefont{H.}~\bibnamefont{Krivine}},
  \bibinfo{author}{\bibfnamefont{J.}~\bibnamefont{Treiner}}, \bibnamefont{and}
  \bibinfo{author}{\bibfnamefont{O.}~\bibnamefont{Bohigas}},
  \bibinfo{journal}{Nuclear Physics A} \textbf{\bibinfo{volume}{336}},
  \bibinfo{pages}{155} (\bibinfo{year}{1980}).

\bibitem[{\citenamefont{Bartel et~al.}(1982)\citenamefont{Bartel, Quentin,
  Brack, Guet, and H{\aa}kansson}}]{bar82}
\bibinfo{author}{\bibfnamefont{J.}~\bibnamefont{Bartel}},
  \bibinfo{author}{\bibfnamefont{P.}~\bibnamefont{Quentin}},
  \bibinfo{author}{\bibfnamefont{M.}~\bibnamefont{Brack}},
  \bibinfo{author}{\bibfnamefont{C.}~\bibnamefont{Guet}}, \bibnamefont{and}
  \bibinfo{author}{\bibfnamefont{H.-B.} \bibnamefont{H{\aa}kansson}},
  \bibinfo{journal}{Nuclear Physics A} \textbf{\bibinfo{volume}{386}},
  \bibinfo{pages}{79} (\bibinfo{year}{1982}).

\bibitem[{\citenamefont{Rashdan}(2000)}]{ras00}
\bibinfo{author}{\bibfnamefont{M.}~\bibnamefont{Rashdan}},
  \bibinfo{journal}{Modern Physics Letters A} \textbf{\bibinfo{volume}{15}},
  \bibinfo{pages}{1287} (\bibinfo{year}{2000}).

\bibitem[{\citenamefont{Goriely et~al.}(2013)\citenamefont{Goriely, Chamel, and
  Pearson}}]{gor13}
\bibinfo{author}{\bibfnamefont{S.}~\bibnamefont{Goriely}},
  \bibinfo{author}{\bibfnamefont{N.}~\bibnamefont{Chamel}}, \bibnamefont{and}
  \bibinfo{author}{\bibfnamefont{J.}~\bibnamefont{Pearson}},
  \bibinfo{journal}{Physical Review C} \textbf{\bibinfo{volume}{88}},
  \bibinfo{pages}{024308} (\bibinfo{year}{2013}).

\bibitem[{\citenamefont{G{\'o}mez et~al.}(1992)\citenamefont{G{\'o}mez, Prieto,
  and Navarro}}]{gom92}
\bibinfo{author}{\bibfnamefont{J.}~\bibnamefont{G{\'o}mez}},
  \bibinfo{author}{\bibfnamefont{C.}~\bibnamefont{Prieto}}, \bibnamefont{and}
  \bibinfo{author}{\bibfnamefont{J.}~\bibnamefont{Navarro}},
  \bibinfo{journal}{Nuclear Physics A} \textbf{\bibinfo{volume}{549}},
  \bibinfo{pages}{125} (\bibinfo{year}{1992}).

\bibitem[{\citenamefont{Sadoudi et~al.}(2013)\citenamefont{Sadoudi, Bender,
  Bennaceur, Davesne, Jodon, and Duguet}}]{sad13}
\bibinfo{author}{\bibfnamefont{J.}~\bibnamefont{Sadoudi}},
  \bibinfo{author}{\bibfnamefont{M.}~\bibnamefont{Bender}},
  \bibinfo{author}{\bibfnamefont{K.}~\bibnamefont{Bennaceur}},
  \bibinfo{author}{\bibfnamefont{D.}~\bibnamefont{Davesne}},
  \bibinfo{author}{\bibfnamefont{R.}~\bibnamefont{Jodon}}, \bibnamefont{and}
  \bibinfo{author}{\bibfnamefont{T.}~\bibnamefont{Duguet}},
  \bibinfo{journal}{Physica Scripta} \textbf{\bibinfo{volume}{2013}},
  \bibinfo{pages}{014013} (\bibinfo{year}{2013}).

\bibitem[{\citenamefont{Onsi et~al.}(1994)\citenamefont{Onsi, Przysiezniak, and
  Pearson}}]{onsi1994equation}
\bibinfo{author}{\bibfnamefont{M.}~\bibnamefont{Onsi}},
  \bibinfo{author}{\bibfnamefont{H.}~\bibnamefont{Przysiezniak}},
  \bibnamefont{and} \bibinfo{author}{\bibfnamefont{J.}~\bibnamefont{Pearson}},
  \bibinfo{journal}{Physical Review C} \textbf{\bibinfo{volume}{50}},
  \bibinfo{pages}{460} (\bibinfo{year}{1994}).

\bibitem[{\citenamefont{Ko et~al.}(1974)\citenamefont{Ko, Pauli, Brack, and
  Brown}}]{ko1974microscopic}
\bibinfo{author}{\bibfnamefont{C.}~\bibnamefont{Ko}},
  \bibinfo{author}{\bibfnamefont{H.}~\bibnamefont{Pauli}},
  \bibinfo{author}{\bibfnamefont{M.}~\bibnamefont{Brack}}, \bibnamefont{and}
  \bibinfo{author}{\bibfnamefont{G.}~\bibnamefont{Brown}},
  \bibinfo{journal}{Nuclear Physics A} \textbf{\bibinfo{volume}{236}},
  \bibinfo{pages}{269} (\bibinfo{year}{1974}).

\bibitem[{\citenamefont{Lesinski et~al.}(2006)\citenamefont{Lesinski,
  Bennaceur, Duguet, and Meyer}}]{Lesinski2006}
\bibinfo{author}{\bibfnamefont{T.}~\bibnamefont{Lesinski}},
  \bibinfo{author}{\bibfnamefont{K.}~\bibnamefont{Bennaceur}},
  \bibinfo{author}{\bibfnamefont{T.}~\bibnamefont{Duguet}}, \bibnamefont{and}
  \bibinfo{author}{\bibfnamefont{J.}~\bibnamefont{Meyer}},
  \bibinfo{journal}{Physical Review C} \textbf{\bibinfo{volume}{74}},
  \bibinfo{pages}{044315} (\bibinfo{year}{2006}).

\bibitem[{\citenamefont{Tondeur et~al.}(1984)\citenamefont{Tondeur, Brack,
  Farine, and Pearson}}]{tondeur1984}
\bibinfo{author}{\bibfnamefont{F.}~\bibnamefont{Tondeur}},
  \bibinfo{author}{\bibfnamefont{M.}~\bibnamefont{Brack}},
  \bibinfo{author}{\bibfnamefont{M.}~\bibnamefont{Farine}}, \bibnamefont{and}
  \bibinfo{author}{\bibfnamefont{J.}~\bibnamefont{Pearson}},
  \bibinfo{journal}{Nuclear Physics A} \textbf{\bibinfo{volume}{420}},
  \bibinfo{pages}{297} (\bibinfo{year}{1984}).

\bibitem[{\citenamefont{Agrawal et~al.}(2005)\citenamefont{Agrawal, Shlomo, and
  Au}}]{agr05}
\bibinfo{author}{\bibfnamefont{B.}~\bibnamefont{Agrawal}},
  \bibinfo{author}{\bibfnamefont{S.}~\bibnamefont{Shlomo}}, \bibnamefont{and}
  \bibinfo{author}{\bibfnamefont{V.~K.} \bibnamefont{Au}},
  \bibinfo{journal}{Physical Review C} \textbf{\bibinfo{volume}{72}},
  \bibinfo{pages}{014310} (\bibinfo{year}{2005}).

\bibitem[{\citenamefont{Gambacurta et~al.}(2011)\citenamefont{Gambacurta, Li,
  Colo, Lombardo, Van~Giai, and Zuo}}]{gam11}
\bibinfo{author}{\bibfnamefont{D.}~\bibnamefont{Gambacurta}},
  \bibinfo{author}{\bibfnamefont{L.}~\bibnamefont{Li}},
  \bibinfo{author}{\bibfnamefont{G.}~\bibnamefont{Colo}},
  \bibinfo{author}{\bibfnamefont{U.}~\bibnamefont{Lombardo}},
  \bibinfo{author}{\bibfnamefont{N.}~\bibnamefont{Van~Giai}}, \bibnamefont{and}
  \bibinfo{author}{\bibfnamefont{W.}~\bibnamefont{Zuo}},
  \bibinfo{journal}{Physical Review C} \textbf{\bibinfo{volume}{84}},
  \bibinfo{pages}{024301} (\bibinfo{year}{2011}).

\bibitem[{\citenamefont{Brown}(1998)}]{brown1998new}
\bibinfo{author}{\bibfnamefont{B.~A.} \bibnamefont{Brown}},
  \bibinfo{journal}{Physical Review C} \textbf{\bibinfo{volume}{58}},
  \bibinfo{pages}{220} (\bibinfo{year}{1998}).

\bibitem[{\citenamefont{Tondeur et~al.}(2000)\citenamefont{Tondeur, Goriely,
  Pearson, and Onsi}}]{ton00}
\bibinfo{author}{\bibfnamefont{F.}~\bibnamefont{Tondeur}},
  \bibinfo{author}{\bibfnamefont{S.}~\bibnamefont{Goriely}},
  \bibinfo{author}{\bibfnamefont{J.}~\bibnamefont{Pearson}}, \bibnamefont{and}
  \bibinfo{author}{\bibfnamefont{M.}~\bibnamefont{Onsi}},
  \bibinfo{journal}{Physical Review C} \textbf{\bibinfo{volume}{62}},
  \bibinfo{pages}{024308} (\bibinfo{year}{2000}).

\bibitem[{\citenamefont{Chabanat et~al.}(1997)\citenamefont{Chabanat, Bonche,
  Haensel, Meyer, and Schaeffer}}]{cha97}
\bibinfo{author}{\bibfnamefont{E.}~\bibnamefont{Chabanat}},
  \bibinfo{author}{\bibfnamefont{P.}~\bibnamefont{Bonche}},
  \bibinfo{author}{\bibfnamefont{P.}~\bibnamefont{Haensel}},
  \bibinfo{author}{\bibfnamefont{J.}~\bibnamefont{Meyer}}, \bibnamefont{and}
  \bibinfo{author}{\bibfnamefont{R.}~\bibnamefont{Schaeffer}},
  \bibinfo{journal}{Nuclear Physics A} \textbf{\bibinfo{volume}{627}},
  \bibinfo{pages}{710} (\bibinfo{year}{1997}).

\bibitem[{\citenamefont{Steiner et~al.}(2005)\citenamefont{Steiner, Prakash,
  Lattimer, and Ellis}}]{ste05}
\bibinfo{author}{\bibfnamefont{A.~W.} \bibnamefont{Steiner}},
  \bibinfo{author}{\bibfnamefont{M.}~\bibnamefont{Prakash}},
  \bibinfo{author}{\bibfnamefont{J.~M.} \bibnamefont{Lattimer}},
  \bibnamefont{and} \bibinfo{author}{\bibfnamefont{P.~J.} \bibnamefont{Ellis}},
  \bibinfo{journal}{Physics reports} \textbf{\bibinfo{volume}{411}},
  \bibinfo{pages}{325} (\bibinfo{year}{2005}).

\bibitem[{\citenamefont{Rayet et~al.}(1982)\citenamefont{Rayet, Arnould,
  Paulus, and Tondeur}}]{ray82}
\bibinfo{author}{\bibfnamefont{M.}~\bibnamefont{Rayet}},
  \bibinfo{author}{\bibfnamefont{M.}~\bibnamefont{Arnould}},
  \bibinfo{author}{\bibfnamefont{G.}~\bibnamefont{Paulus}}, \bibnamefont{and}
  \bibinfo{author}{\bibfnamefont{F.}~\bibnamefont{Tondeur}},
  \bibinfo{journal}{Astronomy and Astrophysics} \textbf{\bibinfo{volume}{116}},
  \bibinfo{pages}{183} (\bibinfo{year}{1982}).

\bibitem[{\citenamefont{Beiner et~al.}(1975)\citenamefont{Beiner, Flocard,
  Van~Giai, and Quentin}}]{bei75}
\bibinfo{author}{\bibfnamefont{M.}~\bibnamefont{Beiner}},
  \bibinfo{author}{\bibfnamefont{H.}~\bibnamefont{Flocard}},
  \bibinfo{author}{\bibfnamefont{N.}~\bibnamefont{Van~Giai}}, \bibnamefont{and}
  \bibinfo{author}{\bibfnamefont{P.}~\bibnamefont{Quentin}},
  \bibinfo{journal}{Nuclear Physics A} \textbf{\bibinfo{volume}{238}},
  \bibinfo{pages}{29} (\bibinfo{year}{1975}).

\bibitem[{\citenamefont{Pearson and Goriely}(2001)}]{pearson2001isovector}
\bibinfo{author}{\bibfnamefont{J.}~\bibnamefont{Pearson}} \bibnamefont{and}
  \bibinfo{author}{\bibfnamefont{S.}~\bibnamefont{Goriely}},
  \bibinfo{journal}{Physical Review C} \textbf{\bibinfo{volume}{64}},
  \bibinfo{pages}{027301} (\bibinfo{year}{2001}).

\bibitem[{\citenamefont{Van~Giai and Sagawa}(1981)}]{van81}
\bibinfo{author}{\bibfnamefont{N.}~\bibnamefont{Van~Giai}} \bibnamefont{and}
  \bibinfo{author}{\bibfnamefont{H.}~\bibnamefont{Sagawa}},
  \bibinfo{journal}{Physics Letters B} \textbf{\bibinfo{volume}{106}},
  \bibinfo{pages}{379} (\bibinfo{year}{1981}).

\bibitem[{\citenamefont{Goriely et~al.}(2009)\citenamefont{Goriely, Hilaire,
  Girod, and P{\'e}ru}}]{gor09}
\bibinfo{author}{\bibfnamefont{S.}~\bibnamefont{Goriely}},
  \bibinfo{author}{\bibfnamefont{S.}~\bibnamefont{Hilaire}},
  \bibinfo{author}{\bibfnamefont{M.}~\bibnamefont{Girod}}, \bibnamefont{and}
  \bibinfo{author}{\bibfnamefont{S.}~\bibnamefont{P{\'e}ru}},
  \bibinfo{journal}{Physical review letters} \textbf{\bibinfo{volume}{102}},
  \bibinfo{pages}{242501} (\bibinfo{year}{2009}).

\bibitem[{\citenamefont{Vautherin}(1972)}]{vau72}
\bibinfo{author}{\bibfnamefont{D.}~\bibnamefont{Vautherin}},
  \bibinfo{journal}{Phys. Rev. C} \textbf{\bibinfo{volume}{5}},
  \bibinfo{pages}{626} (\bibinfo{year}{1972}).

\bibitem[{\citenamefont{Giannoni and Quentin}(1980)}]{gia80}
\bibinfo{author}{\bibfnamefont{M.}~\bibnamefont{Giannoni}} \bibnamefont{and}
  \bibinfo{author}{\bibfnamefont{P.}~\bibnamefont{Quentin}},
  \bibinfo{journal}{Physical Review C} \textbf{\bibinfo{volume}{21}},
  \bibinfo{pages}{2076} (\bibinfo{year}{1980}).

\bibitem[{\citenamefont{Lalazissis et~al.}(2005)\citenamefont{Lalazissis,
  Nik{\v{s}}i{\'c}, Vretenar, and Ring}}]{lal05}
\bibinfo{author}{\bibfnamefont{G.}~\bibnamefont{Lalazissis}},
  \bibinfo{author}{\bibfnamefont{T.}~\bibnamefont{Nik{\v{s}}i{\'c}}},
  \bibinfo{author}{\bibfnamefont{D.}~\bibnamefont{Vretenar}}, \bibnamefont{and}
  \bibinfo{author}{\bibfnamefont{P.}~\bibnamefont{Ring}},
  \bibinfo{journal}{Physical Review C} \textbf{\bibinfo{volume}{71}},
  \bibinfo{pages}{024312} (\bibinfo{year}{2005}).

\bibitem[{\citenamefont{Becker et~al.}(2017)\citenamefont{Becker, Davesne,
  Meyer, Navarro, and Pastore}}]{bec17}
\bibinfo{author}{\bibfnamefont{P.}~\bibnamefont{Becker}},
  \bibinfo{author}{\bibfnamefont{D.}~\bibnamefont{Davesne}},
  \bibinfo{author}{\bibfnamefont{J.}~\bibnamefont{Meyer}},
  \bibinfo{author}{\bibfnamefont{J.}~\bibnamefont{Navarro}}, \bibnamefont{and}
  \bibinfo{author}{\bibfnamefont{A.}~\bibnamefont{Pastore}},
  \bibinfo{journal}{Physical Review C} \textbf{\bibinfo{volume}{96}},
  \bibinfo{pages}{044330} (\bibinfo{year}{2017}).

\end{thebibliography}

\end{document}